\documentclass[10pt]{IEEEtran}
\usepackage{subfigure}
\usepackage{caption}
\usepackage{amsmath,amsthm,amssymb,color,url,multirow}
\usepackage[pdftex]{graphicx}
\usepackage{tikz}
\usepackage{cite}

\newtheorem{rem}{Remark}

\def\p0{{\pmb 0}}

\long\def\comment#1{}

\newfont{\bbb}{msbm10 scaled 700}

\newfont{\bbc}{msbm10 scaled 1100}
\newcommand{\CC}{\mbox{\bbc C}}

\newcommand{\RR}{\mbox{\bbc R}}

\newcommand{\EE}{\mbox{\bbc E}}

\newcommand{\av}{{\pmb a}}

\newcommand{\dv}{{\pmb d}}
\newcommand{\ev}{{\pmb e}}

\newcommand{\hv}{{\pmb h}}

\newcommand{\wv}{{\pmb w}}
\newcommand{\xv}{{\pmb x}}
\newcommand{\yv}{{\pmb y}}
\newcommand{\zv}{{\pmb z}}
\newcommand{\zerov}{{\pmb 0}}

\newcommand{\Bm}{{\pmb B}}

\newcommand{\Em}{{\pmb E}}

\newcommand{\Gm}{{\pmb G}}
\newcommand{\Hm}{{\pmb H}}
\newcommand{\Id}{{\pmb I}}

\newcommand{\Pm}{{\pmb P}}

\newcommand{\Rm}{{\pmb R}}

\newcommand{\Um}{{\pmb U}}
\newcommand{\Vm}{{\pmb V}}

\newcommand{\Xm}{{\pmb X}}

\newcommand{\Bc}{{\cal B}}
\newcommand{\Cc}{{\cal C}}

\newcommand{\Ec}{{\cal E}}
\newcommand{\Fc}{{\cal F}}

\newcommand{\Ic}{{\cal I}}
\newcommand{\Jc}{{\cal J}}
\newcommand{\Kc}{{\cal K}}

\newcommand{\Nc}{{\cal N}}

\newcommand{\Sc}{{\cal S}}

\newcommand{\Wc}{{\cal W}}
\newcommand{\Xc}{{\cal X}}
\newcommand{\Yc}{{\cal Y}}

\newcommand{\muv}{\hbox{\boldmath$\mu$}}

\newcommand{\Lambdam}{\hbox{\boldmath$\Lambda$}}

\newcommand{\Sigmam}{\hbox{\boldmath$\Sigma$}}

\newcommand{\diag}{{\hbox{diag}}}

\renewcommand{\arg}{{\hbox{arg}}}

\newcommand{\herm}{{\sf H}}

\newcommand{\transp}{{\sf T}}

\IEEEoverridecommandlockouts
\renewcommand{\arg}{{\rm arg}}

\begin{document}

\title{Joint Spatial Division and Multiplexing for mm-Wave Channels}

%

\author{\authorblockN{Ansuman Adhikary\authorrefmark{1},
Ebrahim Al Safadi\authorrefmark{1},  Mathew K. Samimi\authorrefmark{2}, Rui Wang\authorrefmark{1},\\
Giuseppe Caire\authorrefmark{1}, Theodore S. Rappaport\authorrefmark{2} and Andreas F. Molisch\authorrefmark{1}}
\thanks{

$^*$ The authors are with the Ming-Hsieh Department of Electrical Engineering, University of Southern California, CA.

$^\dagger$ The authors are with NYU WIRELESS Research Center, ECE Dept, NYU Polytechnic School of Engineering, NY.

This work was partially supported by the collaborative project ``Higher, Denser, Wilder: 5th Generation Wireless Communications'', sponsored by a gift of Intel Labs University Research Office.
The work of A. Adikhary and G. Caire was also partially supported by
ETRI - Electronics and Telecommunications Research Institute, Daejeon, Korea. }}

\maketitle

\begin{abstract}

Massive MIMO systems are well-suited for mm-Wave communications, as large arrays can be built with reasonable form factors, and the
high array gains enable reasonable coverage even for outdoor communications. One of the main obstacles for using such systems in frequency-division duplex mode, namely
the high overhead for the feedback of channel state information (CSI) to the transmitter, can be mitigated by the recently proposed JSDM (Joint Spatial Division and Multiplexing) algorithm. In this paper we analyze the performance of this algorithm in some {\em realistic} propagation channels that take into account the partial overlap of the angular spectra from different users, as well as the sparsity of mm-Wave channels. We formulate the problem of user grouping for two different objectives, namely maximizing spatial multiplexing, and maximizing total received power, in a graph-theoretic framework. As the resulting problems are numerically difficult, we proposed (sub optimum) greedy algorithms as efficient solution methods. Numerical examples show that the different algorithms may be superior in different settings. We furthermore develop a new, ``degenerate'' version of JSDM that only requires average CSI at the transmitter, and thus greatly reduces the computational burden. Evaluations in propagation channels obtained from ray tracing results, as well as in {\em measured} outdoor channels show that this low-complexity version performs surprisingly well in mm-Wave channels.

\end{abstract}

\begin{IEEEkeywords}
5G systems, mm-Waves, MU-MIMO, downlink beamforming, directional channel models, JSDM.
\end{IEEEkeywords}

\section{Introduction}  \label{intro}

Massive MIMO (multiple-input multiple-output) systems are equipped with a large number  (dozens or hundreds) of antenna elements at the
base station (BS) \cite{marzetta2010noncooperative,rusek2013scaling}.
They are intended to be employed in a multi-user MIMO (MU-MIMO) setting,
such that the number of BS antenna elements is much larger than the number of users.
Such an arrangement leads not only to very high spectral efficiency, but also to an important simplification of the signal
processing:  in the idealized regime of independent and isotropically distributed channel vectors, in the limit of an infinite number
of BS antennas,  single-user beamforming, specifically conjugate beamforming (i.e., maximum ratio combining in the receive mode, and maximum
ratio transmission for the transmit mode) eliminates inter-user interference.
Furthermore, the transmit power can be drastically reduced, leading to less interference and a lower energy
consumption of the BS. For all these reasons, massive MIMO has received tremendous
attention in the last years \cite{mimo1,mimo2,mimo4,mimo5,mimo6}.

Massive MIMO is especially promising for systems operating at millimeter (mm-) Wave frequencies.
Due to the short wavelength, very large arrays can be created with a reasonable form factor - a 100-element linear
array is only about 50 cm long at a carrier frequency of 30 GHz.
In light of the extremely large bandwidths that are available for commercial use (up to 7 GHz bandwidth in the 60 GHz band, and around 1 GHz at 28 and 38 GHz carrier frequency), massive MIMO systems in the mm-Wave range are ideally suited for high-capacity transmission and thus anticipated to form an important part of {5G} systems.
While the first commercial mm-Wave products are intended for in-home, short-range communications (e.g., for transmission of uncompressed video)
\cite{ieee80211ad}, the potential of mm-Waves for {\em cellular outdoor} has recently been investigated \cite{rappaportmillimeter,VTC2013,azar201328}.
Experiments have shown a coverage range of more than $200$ m even in non line of sight (NLOS) situations
\cite{azar201328}. Such long-range transmissions require high-gain adaptive
antennas - something that massive MIMO implicitly provides.

For the downlink, massive MIMO systems at mm-Wave (or, for that matter, any other) frequencies require
channel state information at the transmitter (CSIT),
for conjugate beamforming as well as for other, more advanced, forms of
MU-MIMO precoding (see \cite{caire2010multiuser} and references therein).  In most existing papers, it has been assumed that this CSIT can be obtained
from the uplink sounding signals, based on the principle of channel reciprocity \cite{marzetta2010noncooperative}.
However, reciprocity only holds (approximately) in Time Division Duplexing (TDD) systems, where the duplexing time is much shorter than
the coherence time of the channel. In Frequency Division Duplexing (FDD) systems, which are widely used in cellular communications, the spacing between uplink and downlink frequency is - for all practical systems - much larger than the coherence bandwidth of the channel \cite{molisch2010wireless}.
Consequently, CSIT has to be provided through feedback - i.e., each user measures its channel vector in the downlink,
and sends it to the BS in (quantized) form. Due to the large number of BS antenna elements, the overhead for this feedback can become
overwhelming, and methods have to be devised for reducing this load.\footnote{TDD might also require feedback because accurate TDD calibration is difficult to achieve in practical hardware implementations. This is the reason why the only existing commercial
standard that considers MU-MIMO downlink, IEEE 802.11ac, also prescribes explicit downlink training and quantized CSIT feedback, even though it uses TDD.}

Joint Spatial Division and Multiplexing (JSDM) is a recent technique proposed in \cite{adhikary2012joint} to achieve massive-MIMO like
gains for FDD systems (or, more generally, for systems that do not make explicit use of channel reciprocity),
with the added advantage of a reduced requirement for CSIT\footnote{ An approach that exploits the same directional structure of the channel covariance matrix used by JSDM, in order to eliminate pilot contamination in a multi-cell massive MIMO setting, was proposed concurrently and independently in \cite{yin2013coordinated}.}. The idea is to partition the user space into groups of users with {\em approximately similar}
covariances,\footnote{Usually caused by the fact that the multi-path components of such users have similar angles at the BS} and split the beamforming into two stages: a first stage consisting of a pre-beamformer that depends only on the
second order statistics, i.e., the covariances of the user channels, and a second stage comprising
a standard MU-MIMO precoder for spatial multiplexing on the effective channel obtained after pre-beamforming.
The instantaneous CSIT of such an effective channel is easier to acquire
thanks to the considerable dimensionality reduction produced by the pre-beamforming stage.
Also,  JSDM lends itself to a {\em hybrid beamforming} implementation, where pre-beamforming (which changes slowly in time)
may be implemented in the analog RF domain, while the MU-MIMO precoding stage is implemented by standard
baseband processing.  This approach allows the use of a very large number of antennas with a limited number
of baseband-to-RF chains; the latter depends on the number of independent  data streams that we wish to send simultaneously to the users. { A major challenge for massive MIMO in the mm-Wave region is the fact that the Doppler shift scales linearly with frequency, and thus the coherence time is an order of magnitude lower than that of comparable microwave systems. Thus, massive MIMO systems at mm-Wave frequencies need to be restricted to low-mobility scenarios. For comparable speeds of motion, for example, at pedestrian speeds (1 m/s), coherence times are of the order of a few ms at mm-Wave frequencies. Since (outdoor) coherence bandwidths of mm-Wave channels are similar to those of microwave channels \cite{rappaportmillimeter,rappaportmm}, the overall challenges of CSI feedback overhead are then comparable to those of higher-mobility (vehicular) microwave massive-MIMO systems. For example, a $30$ GHz channel for a user moving at $1$ m/s has the same coherence time and bandwidth of a $3$ GHz channel for a user moving at $10$ m/s. In this work, we explicitly assume the availability of perfect channel state information for simplicity (wherever required). In reality, devoting a certain amount of resource to the training phase would discount the achievable throughput by a certain factor \cite{adhikary2012joint}.}

The performance of JSDM depends on the type of channel statistics.
Previous analysis was based on the one-cluster (local scattering) model, which means that the BS ``sees'' the incoming multi-path components (MPCs)
under a very constrained angular range.
This allows for an easy division of the users into sets, whose associated MPCs are disjoint in the angular domain, and can thus be
separated by the pre-beamformers. However, this model does not represent many important scenarios.
For example, in urban environments, high-rise buildings or street canyons can act as important ``common clusters'' that create spatially correlated MPCs for many users \cite{Fuhl_et_al_1998}, \cite{asplund2006cost}, \cite{toeltsch2002statistical}.
Another important effect, which becomes particularly relevant at mm-Wave frequencies, is {\em channel sparsity} - in other words, the number of
significant MPCs is much lower than that for a microwave system operating in a similar environment.
The low number of MPCs enables a further reduction of the CSIT that has to be fed back,
and enables a new ``degenerate'' variant of JSDM, proposed in this paper and referred to as {\em Covariance-based JSDM},
that depends on the channel covariance information only. { In fact, it is well known that, as long as the scattering geometry relative to a given user remains unchanged, the fading channel statistics are wide-sense stationary (WSS).
In particular, this means that the channel covariance matrix is time-invariant.
In a typical scattering scenario, even if a user changes its position by several meters,
the channel second order statistics remain unchanged \cite[Chapter 4]{rappaport2002wireless}. Hence, for a user moving at walking speed ($1$ m/s), the channel fading process is ``locally'' WSS over a time horizon of several seconds,
spanning a very large number of symbol time slots (for example, a $20$ MHz OFDM channel has symbol duration of
$4$ $\mu$s, corresponding to $10^6$ symbols over an interval of $4$s, corresponding to a user position displacement of 4m).
We conclude that it is effectively possible to learn very accurately the channel covariance matrix at the
transmitter side, even without requiring very fast CSIT feedback. This makes our scheme particularly interesting for mm-Waves.}

The main goal of this paper is thus to apply the JSDM approach to {\em realistic} propagation channels
inspired, inter alia, by the recent experimental observations of mm-Wave channels in an urban outdoor environment \cite{azar201328}.
Specifically, our contributions are:
\begin{itemize}
\item We identify a new optimization problem related to the application of JSDM to user groups that are coupled by the presence of common scatterers.
In this case, nulling the common MPCs by pre-beamforming creates linearly independent user groups which can be served simultaneously, on the same transmission resource
(Spatial Multiplexing approach). In contrast, allocating the user groups on orthogonal transmission resources allows to use all the MPCs to convey signal energy
to the users (Orthogonalization approach). The ranking of these two approaches in terms of total system throughput depends on the operating SNR.
\item We generalize the common scatterer problem to the case of many users (or user groups) with partial overlapping of their channel angular spectra
(rigorously defined as the Fourier transform of the antenna correlation function, see Section \ref{binsbins}). For this case,
we develop two new algorithms for user grouping and pre-beamforming design. The first algorithm (Section \ref{opt1}) chooses users that fill many angular
directions (i.e., it tends to serve less users with higher beamforming gain). The second algorithm
(Section \ref{opt2}) maximizes the number of users with at least one mutually non-overlapping set of directions (i.e., it tends to serve more users with lower beamforming gain).
\item We propose a new degenerate version of JSDM (Covariance-based JSDM) that provides orthogonalization of the users
based only on the channel second-order statistics, and thus does not need feedback of the instantaneous CSIT. We discuss for which type of channels
such reduced complexity scheme would perform well with respect to full JSDM, and show through numerical experiments that, as intuition suggest,
covariance-based JSDM works well when the number of users is small with respect to the number of BS antennas and
the channels are formed by a few
MPCs with small angular spread. Remarkably,  this is the case expected in a 5G small-cell system
operating at mm-Wave frequencies.
\item We illustrate the performance of the proposed user selection and JSDM schemes through various numerical examples, based on
multiple
clusters of MPCs, and discrete isolated MPCs, obtained from ray tracing in an outdoor campus environment.
\item We also show sample performance results in {\em measured} propagation channels, from a $28$ GHz measurement campaign recently
carried out in New York City \cite{azar201328}.
\end{itemize}
Overall, JSDM with appropriate user selection and, in some relevant cases, also the simple covariance-based JSDM,
appears to be a very attractive approach for the implementation of multiuser MIMO downlink schemes in outdoor,
small to medium range (10 to 200m)  mm-Wave channels.

The remainder of the paper is organized as follows: Section \ref{sec:dd-impulse-response}
discusses the models for propagation channels as relevant for our analysis;
Section  \ref{sec:jsdm} reviews the principle of JSDM and considers its application in single-cluster and multi-cluster channels.
Section \ref{sec:usr-sel} investigates the novel algorithms for user grouping and selection when the angular spectra of the users are partially
overlapping. Section \ref{sec:results} provides simulation results for multi-cluster, ray-tracing-based,
and measured propagation channels. Some concluding remarks are pointed out in Section \ref{sec:conc}.

{\em Notation}: We use boldface capital letters $(\Xm)$ for matrices, boldface small letters for vectors $(\xv)$,
small letters $(x)$ for scalars and $(\Xc)$ calligraphic letters for sets.
$\Xm^T$ and $\Xm^\herm$ denote the transpose and the Hermitian transpose of
$\Xm$, $||\xv||$ denotes the vector 2-norm of $\xv$.
The union, intersection and difference between two sets $\Xc$ and $\Yc$ are respectively denoted by $\Xc \bigcup \Yc$, $\Xc \bigcap \Yc$ and $\Xc \setminus \Yc$.
The Lebesgue measure of a Borel set $\Xc$ is indicated by $|\Xc|$. If $\Nc$ is a discrete set,
$|\Nc|$ indicates its cardinality. The identity matrix is denoted by $\Id$ (when the dimension is clear from the context) or by $\Id_n$ (when pointing out its dimension $n \times n$).
The indicator function of a set $\Bc$ is denoted by $1\{\Bc\}$.
We also use ${\rm Span}(\Xm)$ to denote the linear subspace generated by columns of $\Xm$
and ${\rm Span}^{\perp}(\Xm)$ for the orthogonal complement of ${\rm Span}(\Xm)$. $\xv \sim \Cc\Nc(\muv;\Sigmam)$
indicates that $\xv$ is a complex circularly-symmetric Gaussian vector with mean $\muv$ and covariance matrix $\Sigmam$.

\section{Spatial Chanel Models} \label{sec:dd-impulse-response}

As we are dealing with a MU-MIMO system, a model for a multiuser, multiantenna channel has to be defined.
Generally, MIMO channel models fall into two categories: (i) physical models, and (ii) analytical models \cite{almers2007survey}.
Physical models describe the physical propagation between transmit array and receive array through the ``double-directional impulse response''
$h(t,\tau, \theta,\psi)$, where $t$ is the time at which the channel is excited, $\tau$ is the considered
delay, and $(\theta,\psi)$ are the angles of departure and arrival, respectively \cite{steinbauer2001double}.
It is common to assume that the double-directional impulse response arises as the sum of the contributions from
discrete MPCs, such that
\begin{equation} \label{ch1}
h(t,\tau,\psi, \theta)=\sum\limits_{p=1}^{\bar{N}(t)} \rho_{p} e^{j \phi_{p}}
\delta(\tau-\tau_{p}) \delta(\theta-\theta_{p}) \delta(\psi-\psi_{p}),
\end{equation}
where the number of MPCs $\bar{N}(t)$ may itself be time-varying.
Note that the above description neglects the effect of polarization and can be generalized to include diffuse radiation by considering intervals of angles and/or delays for which we have a continuum of components, each carrying infinitesimal scattered energy (for a more detailed discussion see, e.g., \cite{wyne2004outdoor}).


Double-directional models are the preferred method for MIMO channel modeling because they are independent of the actual antenna
structures, and efficient methods for incorporating realistic large-scale channel variations are available.
However, for theoretical analysis of transmission schemes, analytical models are often preferred.
These models describe the channel transfer function matrix, i.e., a matrix whose $(i,j)$-th entry is the transfer function from the $j$-th transmit to
the $i$-th receive antenna element. The transfer function matrix subsumes the antenna arrays and the actual propagation
channel; it is thus a description including all effects, for example, antenna coupling from transmit antenna connector to receive antenna connector.
Fortunately, analytical models can be easily derived from double-directional models (though not vice versa).
Specializing to the case of interest in this paper, where the MS has an omni-directional antenna, and the BS is equipped with a uniform linear array,
the double directional channel transfer function between a BS antenna element $m$ and the antenna of a user terminal $k$ is given as
\begin{equation}  \label{multipath}
h_{mk}(f) = \sum_p \rho_{kp} e^{j \phi_{kp}} e^{- j2 \pi f \tau_{kp} } e^{-j2\pi D m \sin \theta_{kp}},
\end{equation}
where $f$ denotes the subcarrier frequency, $D \in (0,\frac{1}{2}]$ is the spacing between two antenna elements
normalized by the carrier wavelength. We focus on the frequency-domain representation of the channel matrix because we assume
the use of OFDM \cite{molisch2010wireless}, which is the modulation of choice of modern cellular and WLAN standards \cite{gast2006802}.
Furthermore,  with respect to (\ref{ch1}), in (\ref{multipath}) we have dropped the dependence on $t$ since we make the
usual assumption of {\em block fading}, for which the channel is locally time-invariant over slots comprising
several OFDM symbols. Therefore,  the number of MPCs, denoted by $\bar{N}_k$, may depend on the user index $k$ but
not explicitly on $t$. Note that block fading is implicitly assumed in virtually all existing cellular and WLAN standards, based on
pilot-aided channel estimation and coherent detection. In addition, small cells operating at mm-Wave frequencies are mainly dedicated to high-throughput
nomadic users, for which the channel time variations are typically very slow. For this reason, in this paper we shall assume that
the channel coefficients $h_{mk}(f)$ are known to the user receiver $k$.\footnote{ The knowledge of CSI at the receiver is commonly achieved in any wireless standard implemented today, and it will also be implemented in mm-Wave standards (e.g., 802.11ad). This is necessary for coherent detection, which is enabled by dedicated pilots that go through the downlink beamforming matrix.} In contrast, we shall discuss in great detail the required channel state information at the transmitter (CSIT) for the MU-MIMO downlink schemes proposed in this paper.

The phase $\phi_{kp}$ depends on the number of wavelengths traveled along the $p$-th path, and
even small fluctuations in the transmitter and receiver positions can produce large variations of such phase,
especially at mm-Wave frequencies. Here, we adopt the common assumption \cite{rappaport2002wireless} that the phases
$\{\phi_{kp} : p = 1,\ldots, \bar{N}_k\}$ are uniformly distributed on $[0,2\pi]$ and mutually independent.
This implies uncorrelated scattering \cite{bello1963characterization}, which is a widely accepted assumption in
channel modeling.  In this case, the space-frequency covariance between $h_{mk}(f_1)$ and $h_{nk}(f_2)$, i.e., the covariance between
the channel of antenna element $m$ at frequency $f_1$ and that of antenna element $n$ at frequency $f_2$, is given by
\begin{figure*}
\begin{eqnarray} \label{eqn:ddir-cov-temp}
\EE[h_{mk}(f_1) h_{nk}^*(f_2)] &=& \EE \left[ \sum_p \sum_l \rho_{kp} \rho_{kl}^* e^{j (\phi_{kp} - \phi_{kl})} e^{- j2 \pi (f_1 \tau_{kp} - f_2 \tau_{kl}) } e^{-j2\pi D (m \sin \theta_{kp} - n \sin \theta_{kl})} \right] \nonumber\\
&=&  \sum_p \sum_l \rho_{kp} \rho_{kl}^* \EE \left[ e^{j (\phi_{kp} - \phi_{kl})} \right] e^{- j2 \pi (f_1 \tau_{kp} - f_2 \tau_{kl}) } e^{-j2\pi D (m \sin \theta_{kp} - n \sin \theta_{kl})} \nonumber\\
&=& \sum_p |\rho_{kp}|^2 e^{-j2\pi D (m - n) \sin \theta_{kp}} e^{-j2\pi (f_1 - f_2) \tau_{kp}}.
\end{eqnarray}
\end{figure*}
In particular, we have the well-known result (common to all uncorrelated scattering channel models) that the channel is wide-sense stationary with
respect to frequency, i.e.,  that the channel spatial covariance is independent of the subcarrier $f$, and the covariance for different subcarriers $f_1$ and $f_2$ depends only
on the subcarrier difference $f_1 - f_2$. Furthermore, for uniform linear arrays, we also have that the channel spatial covariance depends
only on the spatial difference  $D(m-n)$ between the antennas. In particular, letting $M$ denote the number of BS antennas,
the $M \times M$ channel spatial covariance of the user channel vector $\hv_k(f) = (h_{1k}(f), \ldots, h_{Mk}(f))^\transp$ is given by
\begin{equation} \label{eqn:ddir-cov}
\Rm_k = \EE[ \hv_k(f) \hv_k^\herm(f) ] = \sum_p  |\rho_{kp}|^2 \av(\theta_{kp}) \av^\herm(\theta_{kp})
\end{equation}
where we define the linear array response for angle of arrival $\theta$ as
\begin{equation}
\av(\theta) = \left [ \begin{array}{c}
1 \\
e^{-j2\pi D \sin \theta} \\
e^{-j2\pi D 2 \sin \theta} \\
\vdots \\
e^{-j2\pi D (M-1) \sin \theta} \end{array} \right ] ~~
\end{equation}

After these general modeling considerations, we now turn to the specific double-directional models occurring most often in practical situations.
It is well-established that the MPCs tend to occur in {\em clusters} in the delay/angle plane, corresponding to interaction with physical
clusters of scatterers\footnote{Strictly speaking, the scatterers should be called ``interacting objects (IOs)'', since the interaction
of the MPCs with the objects might not only be diffuse scattering but also specular reflection or diffraction. However, the name ``scatterers'' for such IOs is widely used in the literature, so that we follow this convention.}
in the real world. The first, simplest, and still most widely used of such clustered models is the ``one-ring'' model \cite{lee1973effects},
in which the scatterers are located on a circle around the MS.\footnote{We use here a slight modification of this model, in which the scatterers are distributed such that the density of scatterers, as seen from the BS, is uniform in a limited angular range. While in \cite{adhikary2012joint} we also called this model ``one-ring'', in this paper, we call it ``one-cluster'' in order to avoid confusion with the original model of \cite{lee1973effects}.}
 However, measurements have shown that this simple model is mostly
applicable in (flat) rural and suburban areas. In metropolitan areas as well as hilly terrains, additional ``far'' scatterer clusters such as high-rise buildings
can occur. While the local clusters ``belong'' to a particular user (see Section \ref{subsec:one-ring}), the far clusters can contribute to the MPCs of many different users
(see Section\ref{subsec:multiple-ring}),
since they are ``visible'' to all of them \cite{asplund2006cost}.
Further clustering can occur in scenarios where wave guiding through street canyons is dominant;
this is especially important if the BS antenna is below rooftop \cite{toeltsch2002statistical}.

An important feature of propagation at mm-Wave frequencies is
a pronounced sparsity of the double-directional impulse response \cite{rappaportmillimeter}. This arises from two major effects: (i) the specular reflection coefficient
at (inevitably) rough house surfaces decreases, while more power is shifted into diffuse components. Consequently, only MPCs that undergo one or two
reflections carry significant power (as opposed to microwaves, which often can have significant power even after 5 or more reflections);
(ii) diffraction becomes less prominent, so that MPCs that propagate ``around a corner'' are suppressed.
Thus, while at microwave frequencies the number of relevant MPCs can easily reach 40 (for each user position), that number is often less than
10 at millimeter waves.

\section{Joint Spatial-Division and Multiplexing} \label{sec:jsdm}

In this section we review the MU-MIMO precoding approach of \cite{adhikary2012joint}, known as Joint Spatial-Division
and Multiplexing (JSDM); note that the main idea was already outlined in Section \ref{intro}.
Consider the downlink of a wireless system formed by a BS equipped with $M$ antennas
and serving $K$ users, each equipped with a single antenna.
We focus on a fixed OFDM subcarrier and drop the frequency variable $f$ for the sake of notation simplicity.

Suppose that the $K$ users are partitioned into $G$ groups, where the $K_g$ users in group $g$ have statistically independent but identically distributed
channels, with a common covariance matrix $\Rm_g=\Um_g\mathbf{\Lambda}_g\Um_g^{\herm}$. Denoting user $k$ in group $g$ by the index $g_k$,
its channel vector is given by $\hv_{g_k}=\Um_g\mathbf{\Lambda}_g^{\frac{1}{2}}\wv_{g_k}$, where $\wv_{g_k} \sim \Cc\Nc(\zerov, \Id_{r_g})$ is an i.i.d. Gaussian vector
(also independent across different users), $\Um_g$ is a tall unitary matrix of dimensions $M \times r_g$, $\Lambdam_g$ is $r_g \times r_g$ diagonal positive definite, and
$r_g$ denotes the rank of $\Rm_g$.  Letting $\Hm_g = [\hv_{g_1},\ldots,\hv_{g_{K_g}}]$ and $\underline{\Hm} = [\Hm_1,\ldots,\Hm_G]$ denote the group $g$
channel matrix and the overall system channel matrix, respectively, the received vector of signals at all the served users is given by
\begin{equation} \label{yy}
\yv = \underline{\Hm}^{\herm} \Vm \dv + \zv.
\end{equation}
{ $\yv \in \CC^{K}$ is the concatenated vector of signals received by the users, $\Vm \in \CC^{M \times K}$ is the precoding matrix, $\dv \in \CC^{K} $ is the vector of transmitted data streams and $\zv \in \CC^{K}$ is Additive White Gaussian Noise with i.i.d. entries of mean zero and variance 1.} JSDM makes use of two-stage MU-MIMO precoding, i.e., the precoding matrix is given by  $\Vm=\Bm\Pm$ where the
pre-beamforming matrix is $\Bm = [\Bm_1, \ldots, \Bm_G]$, with blocks of dimensions $M \times b_g$, respectively,  and
the MU-MIMO  precoding matrix is $\Pm = \diag(\Pm_1, \dots, \Pm_G)$, with diagonal blocks of dimensions $b_g \times K_g$, respectively.\footnote{Restricting $\Pm$
to be in block diagonal form is referred to in  \cite{adhikary2012joint} as ``Per-Group-Processing''. This is not the only option for JSDM, but it is the most attractive one
since it requires significantly reduced instantaneous CSIT with respect to other techniques. In this work we focus exclusively on this approach.}
As anticipated before, $\Bm$ depends only on the second-order statistics $\{ \Um_{g},\Lambda_{g} : g = 1,\ldots, G\}$ of the downlink channels\footnote{ The advantage of implementing pre-beamforming in the analog RF domain is that only $b = \sum_g b_g$ RF chains are needed. The cost of baseband processing and baseband to RF modulation scales with the intermediate dimension $b$, while the number of antennas $M$ can be very large. For example, in today's LTE technology, large tower-mounted base stations have typically $4$ large radiating elements each formed by $16$ couples of dipoles, forming $8$ cross-polarized pairs. These $64$ elements are driven by a fixed beamforming network creating a sector. Hence, they operate as a big fixed phased array, with $4$ input ports and $64$ outputs. Although in today's implementation this array radiates in a fixed pre-determined way, it is expected that in the near future, efficient reconfigurable RF architectures will be implemented at competitive cost, size and energy efficiency \cite{bdlabs}.},
whereas the MU-MIMO precoding matrices $\Pm_g$ are functions of the corresponding instantaneous ``effective" channels
$\underline{\textsf{\Hm}}_g = \Bm_g^\herm \Hm_g$. As a result, (\ref{yy}) can be re-written as
\begin{eqnarray}
\yv & = & \left [ \begin{array}{c} \yv_1 \\ \vdots \\ \yv_G \end{array} \right ] \nonumber\\
& = &\left [ \begin{array}{c}
     \Hm_1^\herm \Bm_1 \Pm_1 \dv_1 + \sum_{g' \neq 1} \Hm_1^\herm \Bm_{g'} \Pm_{g'} \dv_{g'} + \zv_1 \\
     \vdots \\
     \Hm_G^\herm \Bm_G \Pm_G \dv_G + \sum_{g' \neq G} \Hm_G^\herm \Bm_{g'} \Pm_{g'} \dv_{g'} + \zv_G \end{array} \right ]\nonumber\\
\end{eqnarray}
Furthermore, by appropriate group selection and pre-beamforming design, it is possible to exactly or approximately eliminate the inter-group interference by enforcing the
condition
\begin{equation}
\Hm_g^{\herm} \Bm_{g'} \approx \mathbf{0},\quad \textmd{for \, all} \, g'\neq g.
\end{equation}
Equality can be enforced exactly if $\textmd{Span}(\Um_g)\nsubseteq \textmd{Span}(\{\Um_{g'}:g'\neq g\})$
for all $g=1,\ldots,G$. This condition requires per-group spatial multiplexing $K_g$ satisfying:
\begin{equation} \label{multiplexing-g}
\textmd{dim}\left(\textmd{Span}(\Um_g)\cap\textmd{Span}^{\perp}(\{\Um_{g'}:g'\neq g\})\right)\geq K_g.
\end{equation}
When the group ranks $r_g$ are too large and enforcing exact Block Diagonalization (BD) would result in a
too small number of spatial data streams  $K_g$ constrained by (\ref{multiplexing-g}), the pre-beamforming matrix
can be designed according to an {\em approximate BD} approach, by selecting $r_g^{\star}$ dominant eigenmodes\footnote{We refer to $r_g^\star$ as the ``effective rank'' of $\Rm_g$. The notion of dominant eigenmodes is left fuzzy on purpose, since this depends on the amount of inter-group interference that the system can tolerate, and this, in turn,
depends on the operating SNR. As shown in \cite{adhikary2012joint}, choosing $r_g^\star$ appropriately is part of the non-trivial optimization of the
JSDM scheme.}
$\Um_g^{\star}$ for each group $g$, such that $\textmd{Span}(\Um_g^{\star})\nsubseteq \textmd{Span}(\{\Um_{g'}^{\star}:g'\neq g\})$
for all $g=1,\ldots,G$. In this case, the constraint on the group spatial multiplexing $K_g$ is relaxed to
\begin{equation}
\textmd{dim}\left(\textmd{Span}(\Um_g^{\star})\cap\textmd{Span}^{\perp}(\{\Um_{g'}^{\star}:g'\neq g\})\right)\geq K_g,
\end{equation}
although the streams will be affected by some residual interference.

\subsection{Application to the one-cluster model} \label{subsec:one-ring}

Consider again the channel model in (\ref{multipath}) and assume that all paths correspond approximately to the same
delay (i.e., $\tau_{kp} = \tau_k \;\; \forall \; p$) and that the $\bar{N}_k$ paths are divided into $N'_k$
groups
of $N \gg 1$
paths each, such that the paths in the $i$-th cluster have approximately the same angle of arrival $\theta_{kp} = \alpha_{ki}$.
Hence, we can write
\begin{equation}  \label{multipath1}
h_{mk} = \sum_{i=1}^{N'_k}  \left ( \sum_{p=(i-1)N}^{iN-1} \rho_{kp} e^{j \phi_{kp}} \right ) e^{-j2\pi D m \sin \alpha_{ki}}.
\end{equation}
Since $N$ is large, by the Central Limit Theorem \cite{grimmett1992probability} we can assume that
$\left ( \sum_{p=(i-1)N}^{iN-1} \rho_{kp} e^{j \phi_{kp}} \right )$ is complex Gaussian circularly symmetric.
It follows that $\hv_k$ is a zero-mean complex Gaussian vector with given covariance matrix $\Rm_k$.
Going to a diffuse scattering limit, where we assume $N'_k \rightarrow \infty$, with uniform scattering energy
$O(1/N'_k)$ and angles $\alpha_{ki}$ spanning the interval $[\theta_k - \Delta_k, \theta_k + \Delta_k]$, we
arrive at the one-cluster scattering model \cite{lee1973effects} with $(m,n)$ channel covariance elements
\begin{equation}
[\Rm_k]_{m,n} = \frac{1}{2 \Delta_k} \int_{\theta_k - \Delta_k}^{\theta_k + \Delta_k} e^{-j 2 \pi D (m - n) \sin \alpha} d \alpha.
\end{equation}
We briefly outline the
approximate BD approach to design the pre-beamforming matrix. Suppose that
the users are partitioned into $G$ co-located groups, each of which is identified by its own one-cluster scattering channel, i.e.,
all users $g_k$ in group $g$ have the same $\theta_g$ and $\Delta_g$.  Defining
\begin{equation}
\Xi_g=[\Um_1^{\star},\ldots, \Um_{g-1}^{\star},\Um_{g+1}^{\star},\ldots,\Um_G^{\star}],
\end{equation}
of dimensions $M \times \sum_{g'\neq g} r_{g'}^{\star}$
and rank $\sum_{g'\neq g} r_{g'}^{\star}$, and letting $[\Em_g^{(1)},\Em_g^{(0)}]$
denote a system of left eigenvectors of $\Xi_g$, we have that $\textmd{Span}(\Em_g^{(0)})=\textmd{Span}^{\perp}(\{\Um_{g'}^{\star}:g'\neq g\})$.

The pre-beamforming matrix $\Bm_g$ is obtained by concatenating the projection
onto $\textmd{Span}(\Em_g^{(0)})$ along with eigen-beamforming along the dominant eigenmodes
of the covariance matrix of the projected channels of group $g$.
Denoting the covariance matrix of $\hat{\hv}_{g_k}=(\Em_g^{(0)})^\herm\hv_{g_k}$ as
\begin{equation}
\widehat{\Rm}_g = (\Em_g^{(0)})^\herm\Um_g\mathbf{\Lambda}_g\Um_g^\herm \Em_g^{(0)}=\Gm_g\mathbf{\Phi}_g\Gm_g^\herm,
\end{equation}
{ where $\Gm_g$ and $\mathbf{\Phi}_g$ denote the matrix of eigenvectors and eigenvalues of $\widehat{\Rm}_g$,} we obtain
\begin{equation}
\Bm_g=\Em_g^{(0)}\Gm_g^{(1)},
\end{equation}
where $\Gm_g^{(1)}$ contains the dominant $b_g$ eigenmodes of $\widehat{\Rm}_g$.
When $b_g \geq K_g > 1$, in order to harness the spatial multiplexing in each group, we consider the effective channel matrix of group $g$ given by
$\underline{\textsf{\Hm}}_g = \Bm_g^\herm \Hm_g$ and use for each group $g$ the classical zero-forcing MU-MIMO precoding given as
\begin{equation} \label{eqn:zf-pgp}
\Pm_g = \zeta_g^2 \underline{\textsf{\Hm}}_g \left( \underline{\textsf{\Hm}}_g^\herm \underline{\textsf{\Hm}}_g \right)^{-1}
\end{equation}
where $\zeta_g^2$ is a power normalization factor.
Note that the number of data streams $K_g$ that can be spatially multiplexed in group $g$
cannot be larger  than the rank of the equivalent channel, given by $b_g$.

\subsection{Multiple scattering clusters} \label{subsec:multiple-ring}

\begin{figure}
  \centering
  \includegraphics[width=8cm]{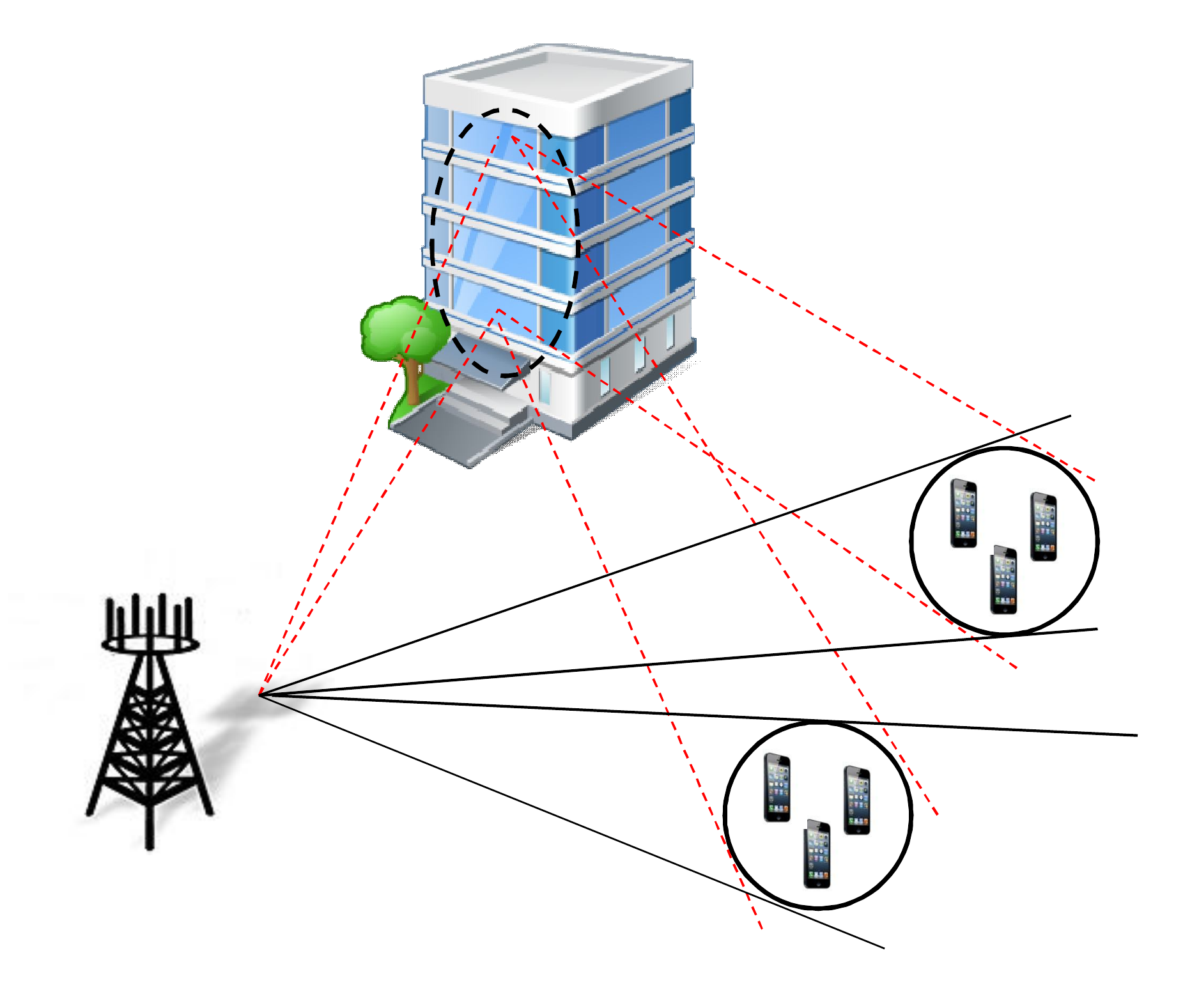}\\
  \caption{Two user groups with local one-cluster scattering and a common scatterer that couples them.}
  \label{fig:common-scat-layout}
\end{figure}

JSDM was originally proposed for a system where users can be partitioned in groups with (approximately) same covariance
subspaces \cite{adhikary2012joint}. Efficient user grouping algorithms  for JSDM are proposed in \cite{adhikary2013joint}.
In any case, the underlying assumption is that the channel vectors in different groups have dominant covariance subspaces that almost do not overlap, such that BD  or approximate BD can efficiently separate the groups on the basis of the channel second-order statistics only.
In this section, we go one step beyond the one-cluster model and consider the application of JSDM
to a more general channel model where each user group is characterized by multiple scattering clusters,
and where these clusters may significantly overlap  (common scatterers).
We formalize the problem and present algorithms for selecting users and allocating spatial dimensions in
Section \ref{sec:usr-sel}.

Figure \ref{fig:common-scat-layout} shows the case of two user groups, each of which has its own cluster of local scatterers,
which share a common  remote scattering cluster. Generalizing this idea, we consider a model where
each user $k$ is characterized by multiple disjoint clusters of scatterers, spanning angle of arrivals in a union of intervals.
For simplicity, we still assume a uniform power distribution over the planar waves impinging on
the BS antenna. This gives rise to a covariance matrix $\Rm_k$ with elements
\begin{equation} \label{eqn:multiple-ring-cov}
[\Rm_k]_{m,n} = \frac{1}{N_k^{\rm cl}} \sum_{c = 1}^{N_k^{\rm cl}} \frac{1}{2 \Delta_{kc}} \int_{\theta_{kc} - \Delta_{kc}}^{\theta_{kc} + \Delta_{kc}}e^{-j 2 \pi D (m-n) \sin \alpha} d \alpha,
\end{equation}
where $N_k^{\rm cl}$ is the number of scattering  clusters associated to user $k$,
and $\theta_{kc}$ and $\Delta_{kc}$ denote the respective azimuth angle
and angular spread of cluster $c$ of user $k$.
One can incorporate different power levels to the scattering clusters by using a
weighted sum of the terms in (\ref{eqn:multiple-ring-cov}).

{ In order to motivate the general problem of selecting users with multiple scattering clusters and gain insight on the design of suitable algorithms for this purpose, we first consider the example of Figure \ref{fig:common-scat-layout}, which shows the effect of a single common scattering cluster.}
Because of the presence of the common scatterers, in order to simultaneously serve users in different groups we need to project the transmit signal
in the orthogonal subspace of the eigendirections corresponding to the common scatterer. In this way, the pre-beamforming projection is able to decouple the two
groups, such that MU-MIMO precoding in each group is able to achieve some per-group spatial multiplexing.
However, in doing so we preclude the possibility of using the paths going through the common scatterer
to convey signal energy to the MSs. Hence, an alternative approach consists of serving the two groups
on different time-frequency slots (orthogonal transmission resources), but maximize the signal energy transfer to each of the groups by exploiting
all the available MPC combining. Summarizing, we have two possible approaches:
\begin{itemize}
\item {\em Multiplexing}: we employ BD to orthogonalize the groups in the spatial domain via the pre-beamforming
matrix. In this way we eliminate inter-group interference, and we are able to serve the two groups on the same transmission resource.
\item {\em Orthogonalization}: we serve the user groups in different channel transmission resources, and use the pre-beamforming matrix to
transmit over all the channel eigenmodes (including the common scatterers) to each group separately.
\end{itemize}

\begin{figure}[!h]
  \centering
  \includegraphics[width=8cm]{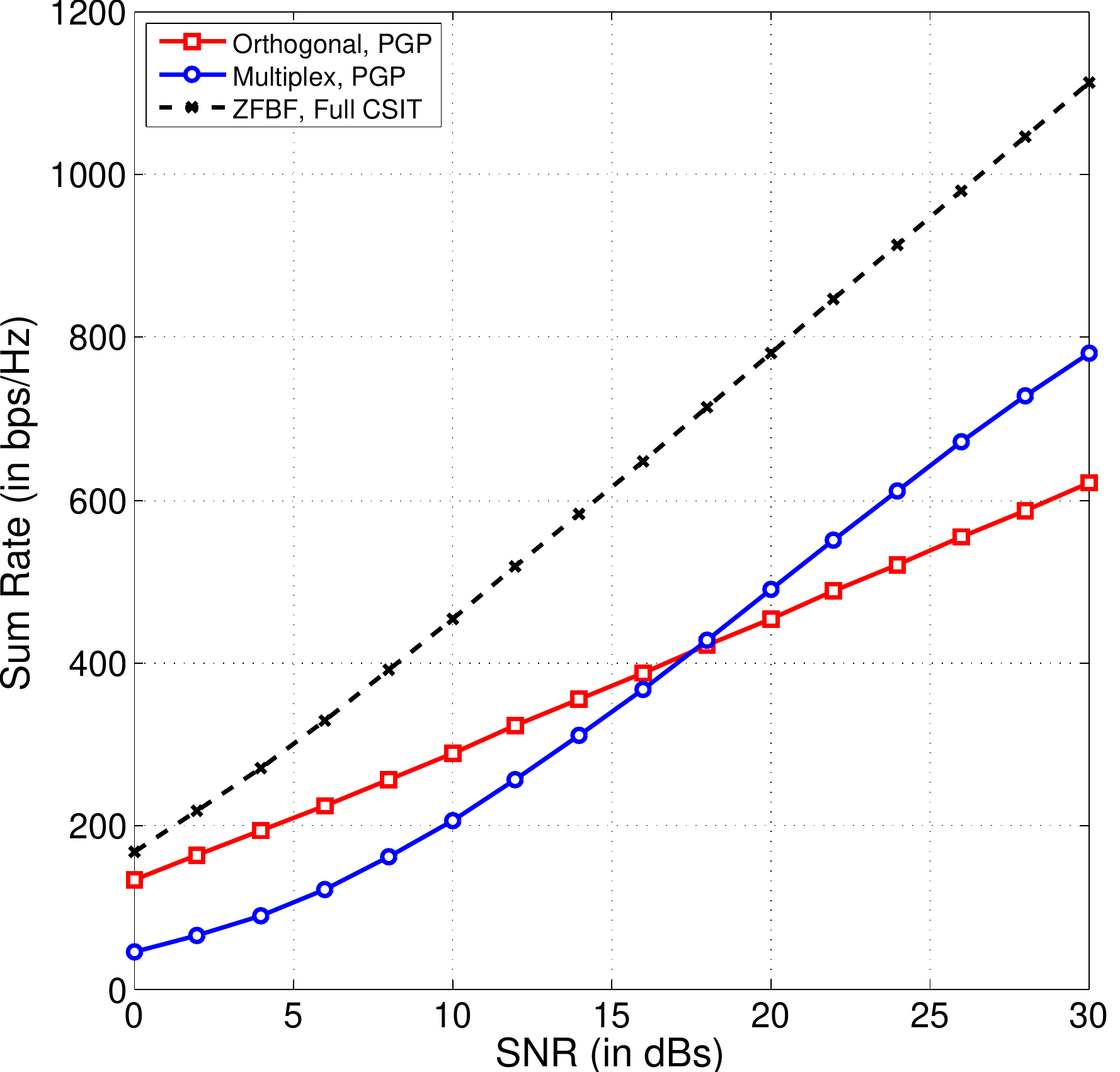}
  \caption{Sum Spectral efficiency (in bits/s/Hz) versus SNR for a scenario with two groups and a common scatterer.}
  \label{fig:common-scat}
\end{figure}

As an example, we set the number of user groups $G = 2$, the total number of users $K = 100$ and the number of BS antennas
$M = 400$. We set the number of users in each group to be equal, i.e., user group $1$ contains $K_1 = 50$ and user group 2 contains $K_2 = 50$ users.
Each of the user groups has two clusters of scatterers, giving $N_1^{\rm cl} = N_2^{\rm cl} = 2$ with one cluster common to both of them (see Figure \ref{fig:common-scat-layout}).
The azimuth angles of the scattering clusters for user group 1 are $\{-45^o,0^o\}$ and those for user group 2 are $\{60^o,0^o\}$. The angular spreads for all the clusters are
taken to be $\Delta = 15^o$. Channel covariances are generated according to (\ref{eqn:multiple-ring-cov}).
The BS power is $P$ and the noise is normalized to 1, giving ${\rm SNR} = P$.
Figure \ref{fig:common-scat} shows the sum spectral efficiency versus SNR for the two approaches mentioned above.
The ``red'' curve corresponds to {\em Orthogonalization} and the ``blue'' curve corresponds to {\em Multiplexing}. { For comparison purposes, we also plot the performance obtained using linear zero forcing beamforming with full channel state information, denoted by the ``black'' curve. It should be noted that for this example, acquiring full CSIT would require $M = 400$ training dimensions (since we are considering an FDD system, and downlink training requirements scale with the number of antennas $M$) in each coherence block. On the other hand, our JSDM scheme requires only $100$ training dimensions (which is a reduction by $4$). This may still be too large for practical scenarios, hence, in the subsequent sections, we propose a degenerate version of JSDM that does not require any instantaneous CSIT.}

We observe that, at low SNR, {\em Orthogonalization} performs better than {\em Multiplexing} due to an increased received power obtained
from the MPCs arising from the common scatterer.
However, at high SNR, {\em Multiplexing} performs much better. This is because even though the received power is less for both groups
after the removal of the common scatterer, more users can be served simultaneously, thereby giving a higher spatial multiplexing,
which is a factor of 2 compared to {\em Orthogonalization} (this is reflected by the slope of the spectral efficiency curves at high SNR).

\section{Application of JSDM to Highly Directional Channels} \label{sec:usr-sel}

In this section, we apply the JSDM approach to highly directional channels as those observed in mm-Wave frequencies.
In particular, we consider the case of channels with multiple scattering clusters, each of which has a different angle of departure and
a narrow angular spread (as in (\ref{eqn:multiple-ring-cov})). In the limit, this reduces to channels formed by discrete and isolated MPCs,
as in the model (\ref{eqn:ddir-cov}). In general, each user (or group of co-located users) has a channel covariance
whose dominant eigenspace ``occupies'' a certain subset of the possible angular directions separable by the BS antenna
array (the resolution of which depends on
$M$ and on the normalized antenna spacing $D$).
Such subsets are formed by unions of disjoint intervals in the angular domain (e.g., see (\ref{eqn:multiple-ring-cov})). Notice here that by assuming intervals, we implicitly consider ``diffuse scattering'' i.e., a continuum of scatterers.
Subsets of different users overlap in some intervals, and are disjoint in other intervals. In fact, this setting is a non-trivial
generalization of the common scatterer problem described in Section  \ref{subsec:multiple-ring}, where in the example we have only two user groups
and three intervals, such that the groups are disjoint on two intervals and overlap on the third, corresponding to the common scatterer.
Thus the general problem that we wish to solve consists of allocating users on the BS spatial dimensions
in order to obtain a good tradeoff between  the spatial multiplexing (number of groups separable by pre-beamforming),
and power gain (which depends on the number of MPCs that are combined to convey signal energy to the receivers).
This problem is combinatorial and can be formulated as an integer program. In order to obtain an efficient and easily computable solution,
we present two integer programming problem formulations and the corresponding greedy user selection algorithms.
As we shall see, each algorithm is suited to a specific scenario, which will be illustrated through
numerical examples in Section \ref{sec:results}.


\subsection{Channel eigenvalue spectrum and angular occupancy}  \label{binsbins}

Using the theory developed in  \cite{adhikary2012joint}, based on Szego's theory of large Toeplitz matrices, the
{\em eigenvalue spectrum} of $\Rm_k$ in the limit of large number of antennas $M$ converges to
the discrete-time Fourier transform of the antenna correlation function, given by $r_k[m - \ell] = [\Rm_k]_{m,\ell}$.
Being a discrete-time Fourier transform of an autocorrelation function,
the eigenvalue spectrum is a function  $\xi_k(f) : \left[-\frac{1}{2}, \frac{1}{2} \right] \longrightarrow \RR^+$.
For the multiple scattering clusters channel model, replicating the derivation
in  \cite{adhikary2012joint} for the one-cluster model, it is immediate to find the eigenvalue spectrum in the form:
\begin{equation}
\xi_k(f) = \left\{ \begin{array}{cc}
\frac{1}{2 N_k^{\rm cl} \Delta_{kc}} \frac{1}{\sqrt{D^2 - f^2}} & f \in \Ic_{kc} \\
0 & f \notin \Ic_{kc} \end{array}\right.\ c \in \{1,2,\ldots,N_k^{\rm cl}\}
\end{equation}
where $\Ic_{kc} = \left( -D \sin (\theta_{kc} + \Delta_{kc}) , -D \sin(\theta_{kc} - \Delta_{kc}) \right)$. In order to handle channels formed by a discrete set of MPCs, we quantize the interval
$[-1/2,1/2]$ into $M$ disjoint intervals (``angular bins'')
of size $\frac{1}{M}$, where bin $\Bc_i$ is centered at $\frac{i}{M} - \frac{1}{2}$ with $i \in \{0,1,\ldots,M-1\}$ and it is wrapped around the interval
$[-1/2,1/2]$ by the periodicity of the discrete-time Fourier transform. We say that a user $k$ ``occupies'' bin $\Bc_i$ if $-D\sin \theta_{kp} \in \Bc_i$.
In addition, we let $\pi(p)$ denote the index of the bin occupied by  the $p$-th MPC.
Then, with a slight abuse of notation, we define $\xi_k(f)$ for the discrete MPC model as the piecewise constant function
\begin{equation}  \label{eqn:ddir-eig-spec}
\xi_k(f) = \sum_{p=1}^{\bar{N}_k} |\rho_{kp}|^2 \cdot 1\{ f \in \Bc_{\pi(p)}\}.
\end{equation}
In both cases, we let  $\Wc_k$ denote the support of $\xi_k(f)$, and define the set function $f_k : \sigma\left (\left[-\frac{1}{2}, \frac{1}{2} \right] \right) \rightarrow \RR_+$ given by
\begin{equation}
f_k(\Xc) = \int_{\Xc} \xi_k(f) df
\end{equation}
where $\Xc$ is an element of the Borel field $\sigma\left (\left[-\frac{1}{2}, \frac{1}{2} \right] \right)$, i.e.,  in particular, it can be any set formed by countable
unions of intervals in $\left[-\frac{1}{2}, \frac{1}{2} \right]$.

In order to formulate the user selection problem\footnote{ The advantage of using linear arrays is the relatively simple mapping between the user angles of departure to the interval $[-1,1]$ (see [9] for details), which gives an elegant mathematical formulation to the user selection problem and enables us to design suitable algorithms. Going beyond a linear array would change the mapping, and the problem needs to be formulated in a different manner.}, we take a graph theoretic approach
and we associate the users to the nodes of a graph, such that node $k$ (corresponding to user $k$) has node weight $\Wc_k$.
An edge $(k,\ell)$ exists in the graph if $\Wc_k \bigcap \Wc_\ell \neq \emptyset$. For such edge, the associated edge weight is
$\Ec_{k \ell} = \Wc_k \cap \Wc_\ell$.

\subsection{Optimization Problem 1}  \label{opt1}

In this case, we aim at maximizing the total ``area'' of the combined eigenvalue spectrum
of the selected users while removing any subspace
overlap between them. The proposed optimization problem takes on the form:

\begin{eqnarray} \label{eqn:optz-1}
{\rm maximize} && \sum_k  f_k\left ( x_k \Wc_k \setminus \left \{ \bigcup_{\ell \in \Nc_k} x_\ell \Ec_{k\ell}  \right \} \right ) \nonumber\\
{\rm subject\ to} && x_k \in \{0,1\}
\end{eqnarray}
with the following notation: for $x \in \{0,1\}$ and $\Wc \in \sigma\left (\left[-\frac{1}{2}, \frac{1}{2} \right] \right)$ we let
$x\Wc = \Wc$ if $x = 1$ and $x\Wc = \emptyset$ if $x = 0$;  $\Nc_k$ denotes the neighborhood of node $k$ in the graph, i.e.,
all the nodes $\ell$ such that an edge $(k,\ell)$ exists.

Note that (\ref{eqn:optz-1}) is an integer optimization problem, whose solution may be computationally complex for real-time implementation,
especially for systems with a large number of users and a large number of angular bins per user channel.  In order to obtain an easily computable
feasible user selection, we resort to a (generally suboptimal) greedy selection algorithm presented below.
For notational simplicity, we denote the objective function of problem (\ref{eqn:optz-1}) by $Q_1(\xv)$, where $\xv = (x_1, \ldots, x_K) \in \{0,1\}^K$.

\paragraph{Greedy Algorithm 1}
\begin{itemize}
\item {\bf Step 1}: Initialize $\xv^{(0)} = \zerov$, the all-zero vector, $Q_1(\xv^{(0)}) = 0$, $\Sc_1 = \emptyset$ and $\Kc = \{1,2,\ldots,K\}$.
\item {\bf Step 2}: For iteration $n$, find an index $k^*$ such that
\begin{equation}
k^* = \arg \max_{k \in \Kc \setminus \Sc_1} Q_1(\xv_k^{(n)}) \nonumber
\end{equation}
where $\xv_k^{(n)} = \xv^{(n)} + \ev_k$, where $\ev_k$ denotes a vector of all zeros except a $1$ in the $k^{\rm th}$ position.
\item {\bf Step 3}: If $Q_1(\xv_{k^*}^{(n)}) > Q_1(\xv^{(n)})$, set $\Sc_1 = \Sc_1 \bigcup \{k^*\}$,
$\xv^{(n+1)} = \xv_{k^*}^{(n)}$,  $n = n + 1$, and go to Step 2.  Else, output $\Sc_1$ as the set of selected users.
\end{itemize}

The greedy algorithm starts by selecting a user that occupies
the maximum area in terms of eigenvalue spectrum
and continues to add more users until
the objective cannot be increased further. From a qualitative perspective, the algorithm implements a form of  {\em Orthogonalization}, by giving preference to users
which occupy a larger area in the eigenvalue spectrum and by penalizing users having a spectral overlap
with the already selected users.

\subsection{Optimization Problem 2}  \label{opt2}
\setcounter{paragraph}{0}

In this case, we wish to maximize the number of served users, provided that they have at least one non-overlapped spectral interval.
The proposed optimization problem takes on the form:
\begin{eqnarray} \label{eqn:optz-2}
{\rm maximize} && \sum_k x_k \nonumber\\
{\rm subject\ to} && x_k \in \{0,1\} \nonumber\\
&& \left[ x_k \Wc_k \setminus \left \{ \bigcup_{\ell \in \Nc_k} x_\ell \Ec_{k\ell} \right \} \right] \nonumber\\
&& \bigcup \left[ \bigcup_{\ell \in \Nc_k} (1 - x_k) \Ec_{k\ell}  \right] \neq \emptyset \ \ \forall \ k
\end{eqnarray}
and $\Nc_k$ denotes all the nodes connected to node $k$. The constraint guarantees that the scheduled user nodes always have one non-overlapping interval, which is non-empty. For the non-scheduled user node, the constraint reduces to a union of edge weights corresponding to its neighbors, which is trivially non-empty (assuming that the graph is connected).

Qualitatively, the optimization problem (\ref{eqn:optz-2}) aims at maximizing the {\em Spatial Multiplexing},
while removing any region of overlap in the angular spectrum of the users.
The solution corresponds to the maximum number of users that can be simultaneously served without
any common region of overlap. Again, since (\ref{eqn:optz-2}) is an integer program, we resort to a (suboptimal) low complexity greedy selection method
that keeps adding users until the feasibility conditions in (\ref{eqn:optz-2}) are satisfied.

\paragraph{Greedy Algorithm 2}
\begin{itemize}
\item {\bf Step 1}: Initialize $\Sc_2 = \emptyset$, $\Kc = \{1,2,\ldots,K\}$ and fix $\epsilon > 0$.
\item {\bf Step 2}: Construct a set $\Fc$ containing all nodes in $\Kc \setminus \Sc_2$ that satisfy the feasibility condition when all nodes in $\Sc_2$ are active, i.e.,
\begin{eqnarray}
\Fc &=& \left \{ k : k \in \Kc \setminus \Sc_2, \left | \Jc_m \right | \geq \epsilon, \ \forall m \in \Sc_2 \cup \{k\}  \right \} \nonumber\\
&& \Jc_m = \Wc_m \setminus \left \{ \bigcup_{\substack{\ell \in \Nc_m \\ \ell \in \Sc_2 \cup \{k\}}} x_\ell \Ec_{m\ell} \right \}
\end{eqnarray}
If $\Fc = \emptyset$, go to Step 5, else go to Step 3.
\item {\bf Step 3}: Find an index $k^* \in \Fc$ such that
\begin{equation} \label{kstar}
k^* = \arg \min_{k \in \Fc} | \Nc_{k} |
\end{equation}
\item {\bf Step 4}: $\Sc_2 = \Sc_2 \cup \{ k^* \}$. Go to Step 2.
\item {\bf Step 5}: Output $\Sc_2$ as the set of selected users.
\end{itemize}
The selection of $k^*$ in (\ref{kstar}) is driven by the heuristic of choosing
a feasible node with minimum number of edges. One can use different heuristics yielding possibly different results.
Finally, $\epsilon$ is a tuning parameter that is used to limit the maximum number of users that can be multiplexed together.
The role of $\epsilon$ is to discard users from getting selected in case they have large overlap regions with other users.

{ Note that the complexity of an optimal exhaustive search user selection algorithm for both (\ref{eqn:optz-1}) and (\ref{eqn:optz-2}) is exponential in the number of users $K$, i.e., $O(2^K)$, whereas the greedy user selection algorithms have a linear complexity, i.e., $O(K)$.} A simple example demonstrating the purpose of the optimization problems 1) and 2) and the corresponding greedy algorithms is given next.
Consider $K = 2$, with $\Wc_1 = (-0.1,0.1) \bigcup (0.2,0.25)$ and $\Wc_2 = (-0.1,0.1) \bigcup (-0.4,-0.3)$.
Also, assume the function $f(\Xc)$ for an interval $\Xc$ is given as $f(\Xc) = |\Xc|$,
the size of the interval. Solving (\ref{eqn:optz-1}) gives the solution $[0\ 1]$ and solving (\ref{eqn:optz-2})
gives $[1\ 1]$ as the solution. This means that with Algorithm 1, only user 2 is selected, while with Algorithm 2 both users are selected.

An important point to note here is that when the channels are highly directional, the eigenvalue spectrum reduces to the form (\ref{eqn:ddir-eig-spec}),
and a user can be viewed as occupying a set of bins corresponding to the angles of arrival of the MPCs. In such a scenario,
if the users are located randomly in the network, the greedy algorithm 2 basically tries to schedule users which have at least one
non-overlapping bin, thereby providing a huge spatial multiplexing.

\subsection{Application of JSDM after selection} \label{subsec:jsdm-after-us}

In this subsection, we briefly summarize the application of Joint Spatial Division and Multiplexing after user selection.
We consider the following two different cases.
\begin{enumerate}
\item {\em JSDM with spatial multiplexing}: In this scenario, users come in groups, either by nature or by the application of user
grouping algorithms. The selection algorithms described earlier provide a set of user groups that can be served simultaneously, in the same transmission resource.
We use approximate BD based on the channel covariances of the selected user groups in order to obtain the JSDM pre-beamformers (see Section \ref{sec:jsdm}).
In this way, pre-beamforming spatially separates the groups. Then, within each group, multiple users are served by spatial multiplexing using a zero-forcing MU-MIMO precoder (see (\ref{eqn:zf-pgp})).
\item {\em Covariance-based JSDM}: In this scheme, irrespective of the number of users in a group, we do not perform spatial multiplexing, i.e.,
only one user per group is served. Mathematically, this means that the pre-beamforming matrices $\Bm_g $ for all groups $g \in \{1,2,\ldots,G\}$ have horizontal dimension
$b_g = 1$, i.e., the pre-beamformer reduces to a single column. This approach can be regarded as a {\em degenerate} version of JSDM where the multiplexing
inside each group is  trivial.  Covariance-based JSDM is attractive from the system simplification viewpoint, since it does not require instantaneous
CSIT to compute the MU-MIMO precoders $\{\Pm_g\}$. On the other hand, when a non-trivial spatial multiplexing per group $K_g > 1$ is possible,
the rate achieved by covariance-based JSDM may be significantly less than what could be achieved by full JSDM.
It is important to remark, though, that in some relevant scenarios the throughput achieved by covariance-based JSDM may be comparable to
that of full JSDM. For example, in a small cell system operating at mm-Wave frequencies,
such that the number of users $K$ is not very large,  and each user channel is formed by discrete MPCs that overlap only on a few common
scattering angles, it can be expected that, after the selection algorithm,
each ``group'' is formed indeed by just a single user.  Therefore, there is no need for further spatial multiplexing inside each group.
This will be evident in some numerical experiments presented in Section \ref{sec:results}.
\end{enumerate}

{
\begin{rem}
From (\ref{eqn:ddir-cov-temp}), we have that the channel covariance matrix of a user $k$ at any given frequency $f$ is independent of the delays $\{\tau_{pk}\}$ of the multi-path components, and is constant with respect to the frequency $f$ (see (\ref{eqn:ddir-cov})). Hence, making a narrowband assuption (e.g., focusing on a single subcarrier of an OFDM system), we can treat the channel covariance as a constant with respect to frequency. Since our algorithms depend only on the channel
covariance matrices, they apply identically whether the channel is frequency selective or frequency flat. Of course, the part of the beamforming scheme that depends on the instantaneous effective channel requires CSIT for every coherence band in frequency. In an extreme case of frequency selectivity, this must be estimated over each OFDM subcarrier, while in a normal case (e.g., channels used in LTE) an estimate per channel resource block (12 adjacent subcarriers) would be sufficient.
\end{rem}}

\section{Numerical Results} \label{sec:results}

We present some numerical experiments demonstrating the performance of the algorithms described in Section \ref{sec:usr-sel}.
We run the algorithms for different scenarios in order to point out  interesting insights on the effect of highly directional channels with common scatterers. We present results for the above discussed multi-cluster model, as well as for even more realistic scenarios generated by ray tracing and measurements.
Before presenting the numerical results, in Section \ref{subsec:ray-trace-channel-x}, we describe the ray tracing setup and in Section \ref{subsec:measurement}, we provide details on the measurement setup.

\subsection{Ray tracing channels} \label{subsec:ray-trace-channel-x}

In order to get channel models even more realistic than the multi-cluster model described above,
we simulate the double directional impulse responses described in Section \ref{sec:dd-impulse-response} with the aid of a
commercial ray-tracing tool, Wireless InSite \cite{wiweb}. This ray tracer
provides efficient and accurate predictions of propagation and communication channel characteristics over $50$ MHz to $100$ GHz in complex environments. Specifically, Wireless InSite performs {\em ray launching}, emitting rays (representing plane waves) from the transmit location into all directions, and following each ray as it interacts (reflection, diffraction, transmission) with the objects in the environment; this continues until either the strength of the ray falls below a specified threshold or it has left the area of interest\footnote{ This commercial ray-tracer does not consider the effects of diffuse MPCs, while there are more advanced ray-tracing tools with the addition of models of diffuse MPCs \cite{degli2011analysis}, \cite{mani2012directional}}.

The input to the program is a digital map of the environment (including footprint and height of the buildings and the electromagnetic characteristics of the building materials). Meanwhile, the effects of trees are non-neglibile in mm-Wave system and thus are modeled by Foliage Feature in Wireless InSite. The output is a list of parameters for
the MPCs that is similar to the result of a double directional channel. Each MPC is associated with a path vector that contains the time averaged path power $P_p = \rho_p^2$, propagation delay $\tau_p$, the azimuth angle of departure $\theta_p$ and arrival $\psi_p$. Like all ray tracers, the accuracy of the program is determined
by the accuracy of the environmental data base, the number of rays launched, and the maximum number of interactions taken into account. Simulation results have been compared to measurements in a variety of settings and shown to provide good agreement \cite{wiweb}.

\begin{figure*}[!hrt]
\centering
\includegraphics[width = 5in]{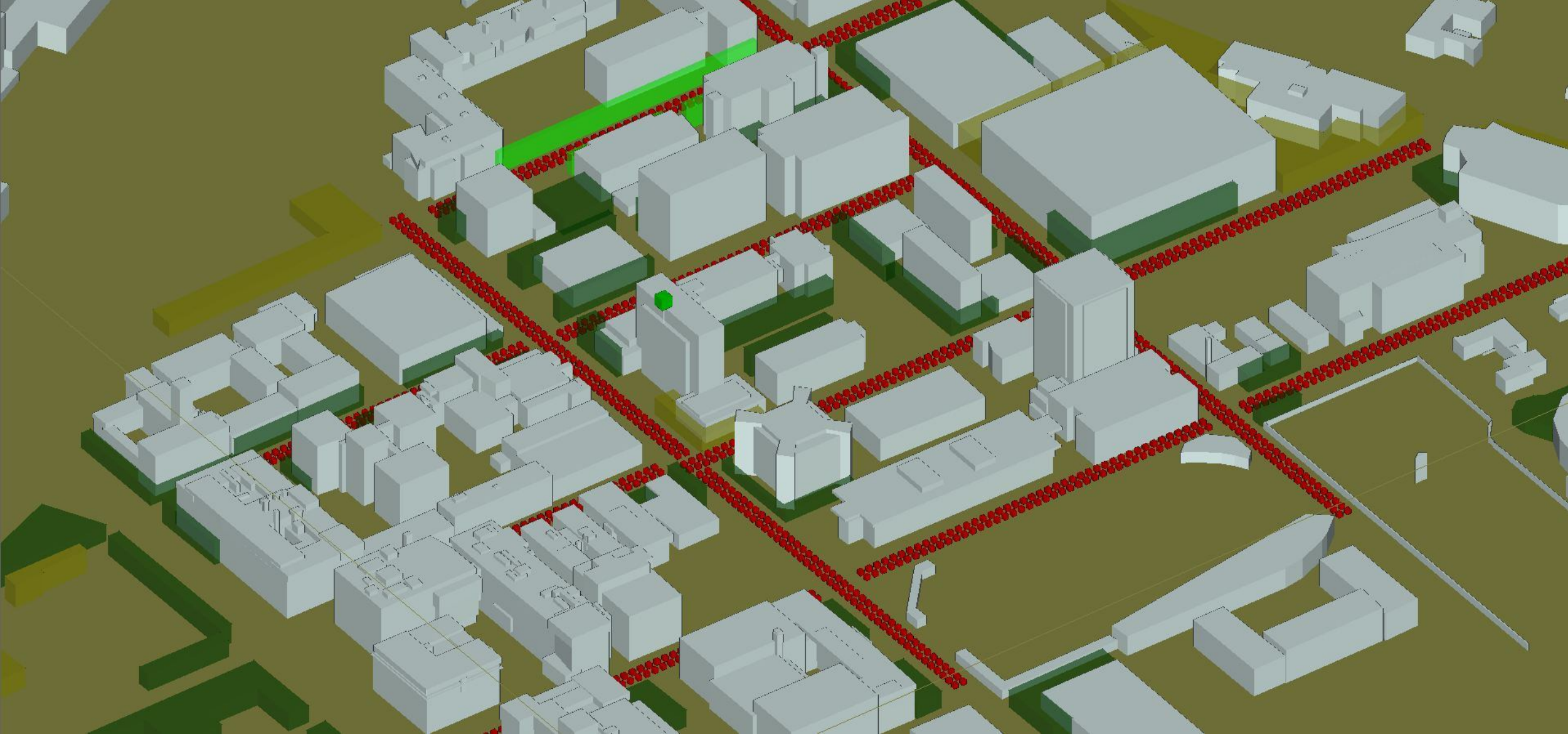}
\caption{Ray-tracing simulation environment}
\label{Fig:Ray-tracing_USC}
\end{figure*}

The simulation has been conducted based on the model of the University of Southern California (USC) main campus,
as shown in Figure \ref{Fig:Ray-tracing_USC}. The green dot is the BS located above the rooftop in the middle of the map,
while simulated MSs are red routes covering all possible streets of the campus. Gray objects represent the buildings, and their building surfaces are modeled with a uniform material for simplicity. The light/dark green 3D polygons denote foliage features with different tree density. In mm-Wave channels, the diffracted MPC will be greatly attenuated, therefore restricting the ray-tracer to consider up to one diffraction is a valid simplification and speeds up the simulation. The detailed simulation configurations are listed in Table \ref{Tab:SimEnvrionPara}.

\begin{table}[!hrt]
\centering
 \begin{tabular}{|l|l|}
 \hline
 {\bf Variable} & {\bf Value} \\
 \hline
 Carrier Frequency  & 28 GHz \\
 \hline
 Antenna Pattern & Isotropic \\
 \hline
 Antenna Polarization & Vertical \\
 \hline
 Tx power & 30 dBm \\
 \hline
 BS height & 45 m \\
 \hline
 MS height & 2 m \\
 \hline
 Maximal Diffraction & 1\\
 \hline
 Maximal Reflection & 10\\
 \hline
 \end{tabular}
 \caption{Ray-tracing simulation parameters of USC campus}
 \label{Tab:SimEnvrionPara}
\end{table}

\subsection{Measured channels} \label{subsec:measurement}

\begin{figure*}
    \centering
 \includegraphics[width=6.5in]{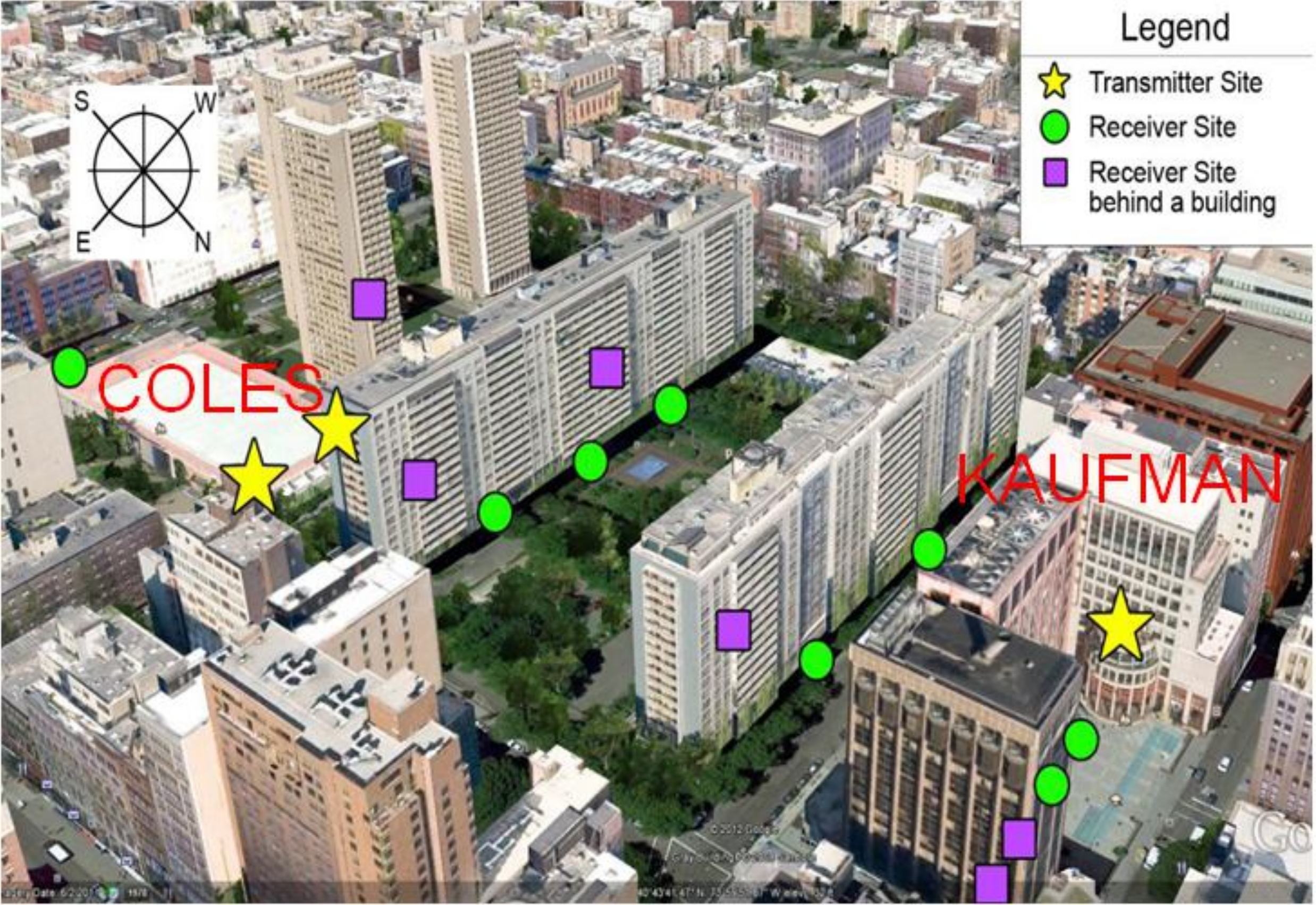}
    \caption{28 GHz cellular measurement locations in Manhattan near the NYU campus. Three base station locations (yellow stars on the one-story rooftop of Coles Recreational Center and five-story balcony of the Kaufman building of Stern Business School) were used to transmit to each of the 25 RX locations within 31 to 423 m. Green dots represent visible RX locations, and purple squares represent RX sites that are blocked by buildings in this image.}
    \label{fig:ManhattanMap}
\end{figure*}

28 GHz wideband propagation measurements of channel impulse responses and received power were made throughout downtown New York City in the summer of 2012. Three different transmitter (BS) locations were selected on NYU buildings, two being on the rooftop of the Coles Sports Center (7 m above ground) and a third on the fifth-floor balcony of the Kaufman Center (17 m above ground). Each transmitter location shared 25 receiver locations with transmitter-receiver separation distances ranging from 31 m to 423 m, for a total of 75 TX-RX distinct RX locations, although only 25 locations with TX-RX separations less than 200 m were able to receive sufficient power for broadband signal capture. Fig.~\ref{fig:ManhattanMap} shows a 3D map of the Manhattan environment where the measurements were performed, and shows the three transmitters (yellow stars) and receiver locations (green dots and purple squares, with green dots representing visible RX locations and purple squares representing RX sites that are blocked by buildings). Typical measurements included:

\begin{itemize}
  \item Line-of-Sight Boresight (LOS-B) $-$ both the TX and RX antennas are pointed directly toward each other (i.e., on boresight) and aligned in both azimuth and elevation angles with a true LOS $-$ no obstructions between the antennas.
  \item Line-of-Sight Non-Boresight (LOS-NB) $-$ both the TX and RX have no obstructions between the antennas, but the antennas are not pointed directly towards each other in azimuth or elevation angles.
  \item Non-Line-of-Sight (NLOS) $-$ the TX and RX have physical obstructions between the antennas. A NLOS environment with moderate obstructions includes trees between TX and RX, or when the RX is slightly behind a building corner. A NLOS environment with heavy obstruction includes the RX completely behind buildings.
\end{itemize}

The measurements were performed using a 800 MHz first zero-crossing RF bandwidth sliding correlator channel sounder with rotational highly directional horn antennas (each with 24.5 dBi gain, or 10$^\circ$ half beamwidth) \cite{rappaportmillimeter,azar201328,zhao13}. The maximum transmitter output power used was 30 dBm, and two highly directional horn antennas of 24.5 dBi (10.9$^\circ$ and 8.6$^\circ$ half-power beamwidths (HPBW) in the azimuth and elevation planes, respectively) were used at the TX and RX, allowing for a total of 178 dB of measurable path loss. The measurement parameters are summarized in Table \ref{tbl:tableSpec}; for further details see  \cite{rappaportmillimeter} and \cite{azar201328}.

\begin{table}

\centering
\begin{tabular}{|l|l|}
\hline
\multicolumn{1}{|l|}{\textbf{Description}} & \textbf{Value}                     \\ \hline
\multirow{2}{*}{Sequence}                    & 11th order PN Code \\
                                           & (Length = 2047) \\ \hline
Transmitted Chip Rate                      & 400 MHz                            \\ \hline
Receiver Chip Rate                         & 399.95 MHz                         \\ \hline
Slide Factor                               & 8000                               \\ \hline
Carrier Frequency                          & 28 GHz                             \\ \hline
NI Digitizer Sampling Rate                 & 2 MSamples/s                    \\ \hline
System measurement range                       & 178 dB                             \\ \hline
Maximum TX Power                           & 30 dBm                             \\ \hline
TX/RX Antenna Gain                         & 24.5 dBi                           \\ \hline
TX/RX Azimuth and Elevation HPBW           & 10.9$^\circ$/8.6$^\circ$                         \\ \hline
TX-RX Synchronization                      & Unsupported                        \\ \hline
\end{tabular}

\caption{28 GHz Channel Sounder Specifications}\label{tbl:tableSpec}

\end{table}

Angle of arrival (AOA) and angle of departure (AOD) measurements were made for every TX-RX location, as described in \cite{rappaportmillimeter}. For our simulations, we use the measurements to produce AOD received power values reflecting measurable signal propagation for all RX locations\footnote{It was observed from mm-Wave field measurements that the power levels of diffuse multipath components in NLOS environments are considerably weaker than those arising from specular reflections. As a result, evaluating our algorithm based on the most significant multipath components does not significantly impact the presented results.}. AOD measurements consisted of rotating the TX antenna in 10$^\circ$ increments in the azimuth plane at a fixed -10$^\circ$ elevation downtilt while the RX antenna remained stationary at fixed elevation and azimuth angles; this fixed direction of the RX antenna was chosen to approximately maximize the received power. 

\subsection{JSDM with spatial multiplexing} \label{subsec:jsdm-mux-mcs}

As stated in Section \ref{subsec:jsdm-after-us}, here we assume that users come in groups, and each group has multiple scattering clusters,
with covariances computed from (\ref{eqn:multiple-ring-cov}).
In order to generate such a scenario, we form a set of non-overlapping scattering clusters
and divide them into two sets. Each cluster of the first set is assigned uniquely to one group, while the clusters of the second set are assigned
randomly to the groups, such that a cluster in the second set may be common to multiple groups. Hence, each user group has its own
scatterer, different from all the other user groups, in addition to some scatterers that are possibly common to other groups.
In our simulations, we generate $10$ scattering clusters at random, and vary the number of user groups $G$ from $2$ to $5$.
The maximum number of scattering clusters for each user group is fixed to $5$.
Within each user group, a finite number of users equal to the rank of the local scattering cluster is assumed.
These users are then spatially multiplexed by ZFBF on the resulting channel obtained after pre-beamforming, which is determined by approximated BD
on the dominant eigenspaces of the selected user groups. We set $M = 400$, and the noise power is normalized to 1, so ${\rm SNR} = P$, where $P$ is the total BS transmission power.

\begin{figure} [!h]
  \centering
  \subfigure[$G = 2$]{
  \includegraphics[width=7cm]{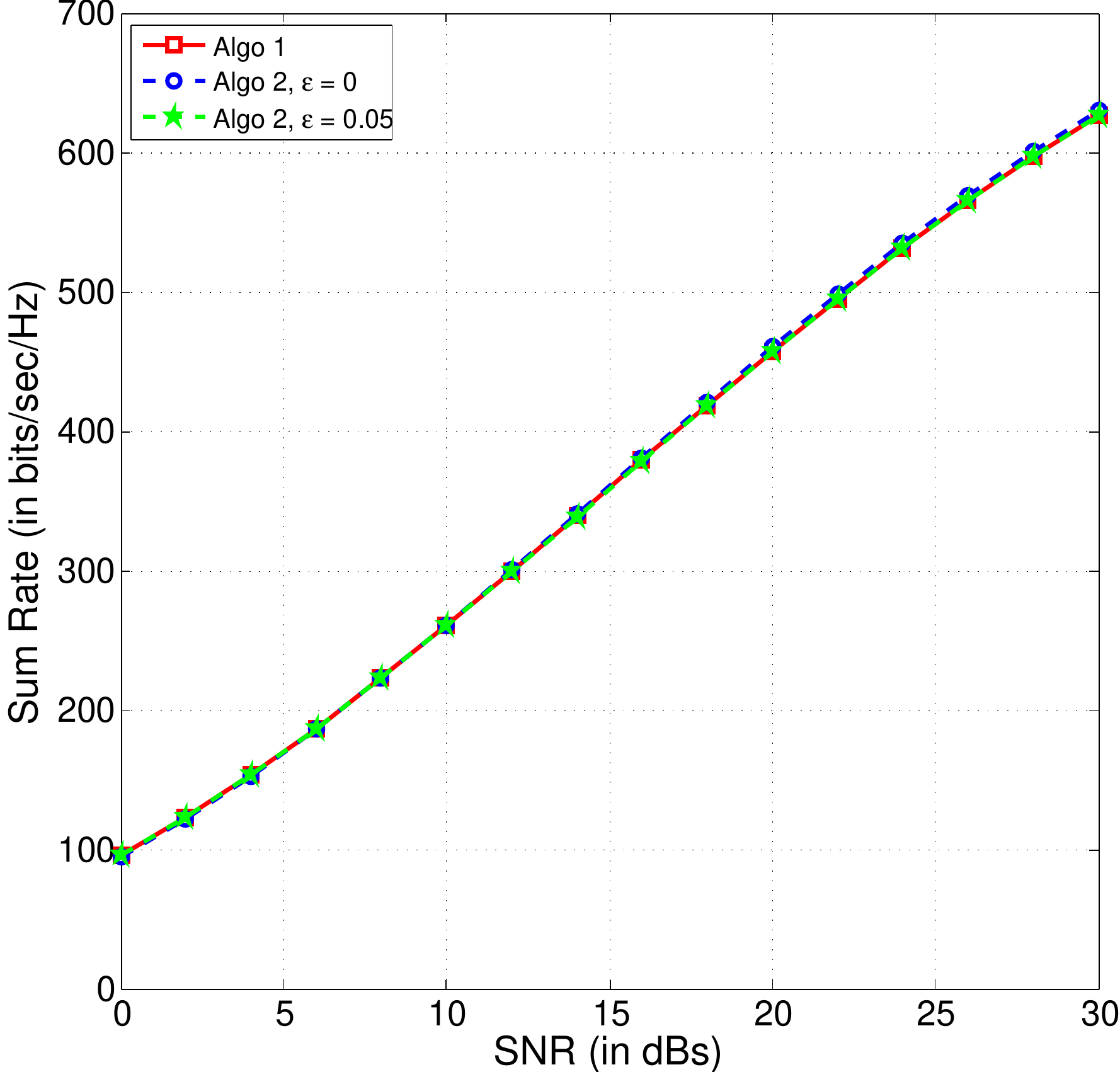}
  \label{fig:rate-mux-mcs-2}
  }
  \subfigure[$G = 5$]{
  \includegraphics[width=7cm]{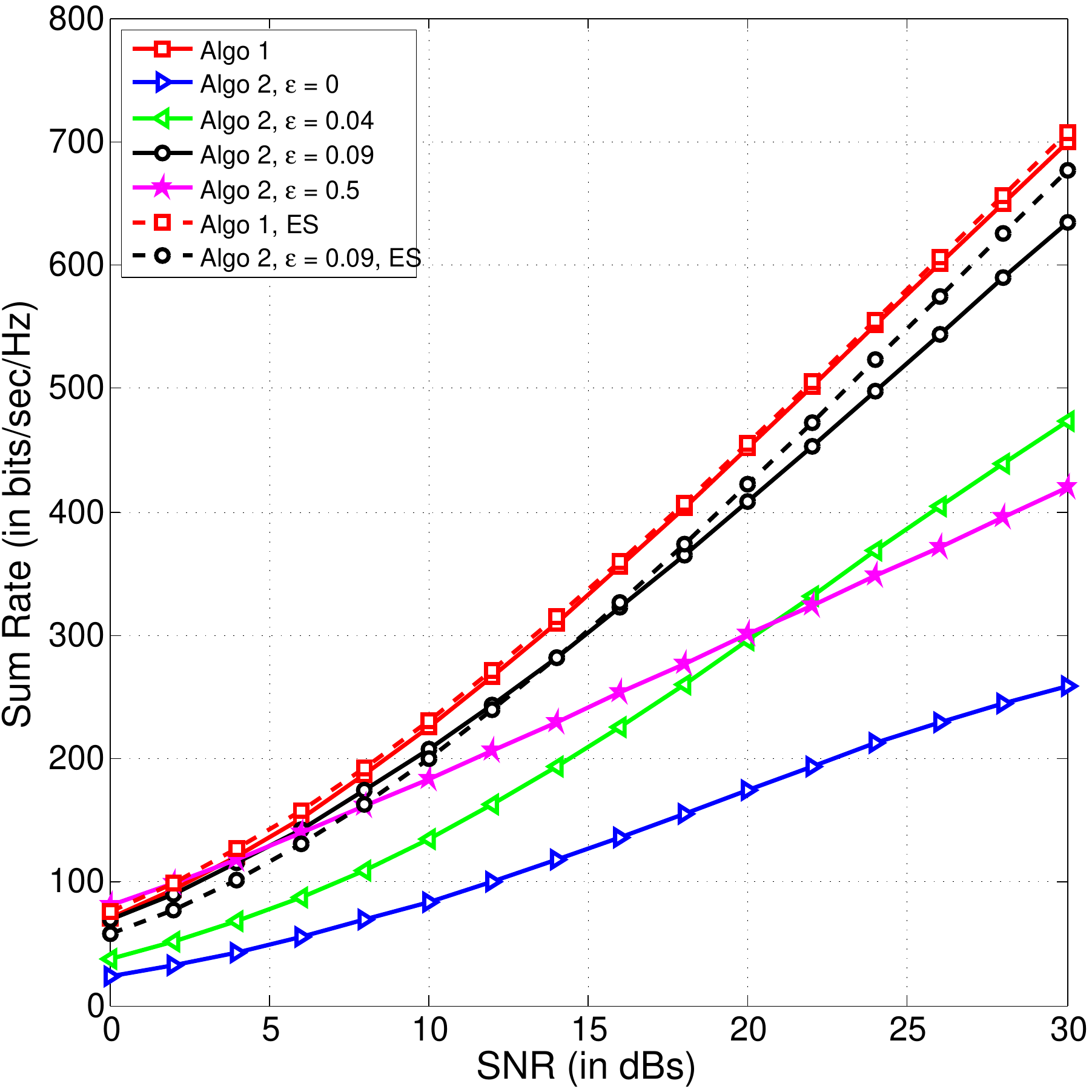}
  \label{fig:rate-mux-mcs-5}
  }
  \caption{Comparison of sum spectral efficiency versus SNR with $G = 2$ and $G = 5$ user groups.
  Each user group has multiple scattering clusters, of which some are common to more than one group.}\label{fig:rate-mux-mcs}
\end{figure}

\begin{figure} [!h]
  \centering
  \subfigure[$G = 2$]{
  \includegraphics[width=7cm]{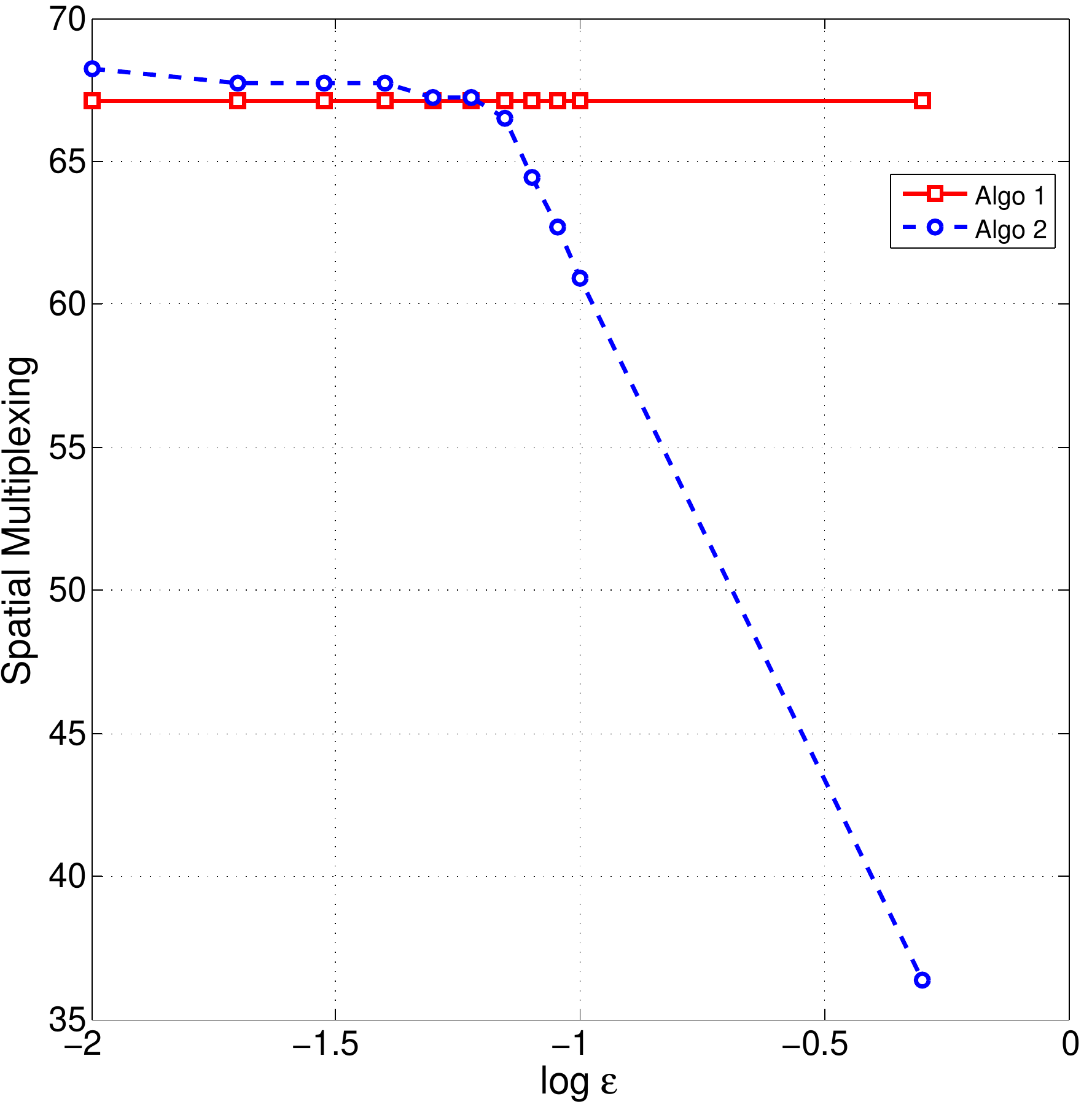}
  \label{fig:user-mux-mcs-2}
  }
  \subfigure[$G = 5$]{
  \includegraphics[width=7cm]{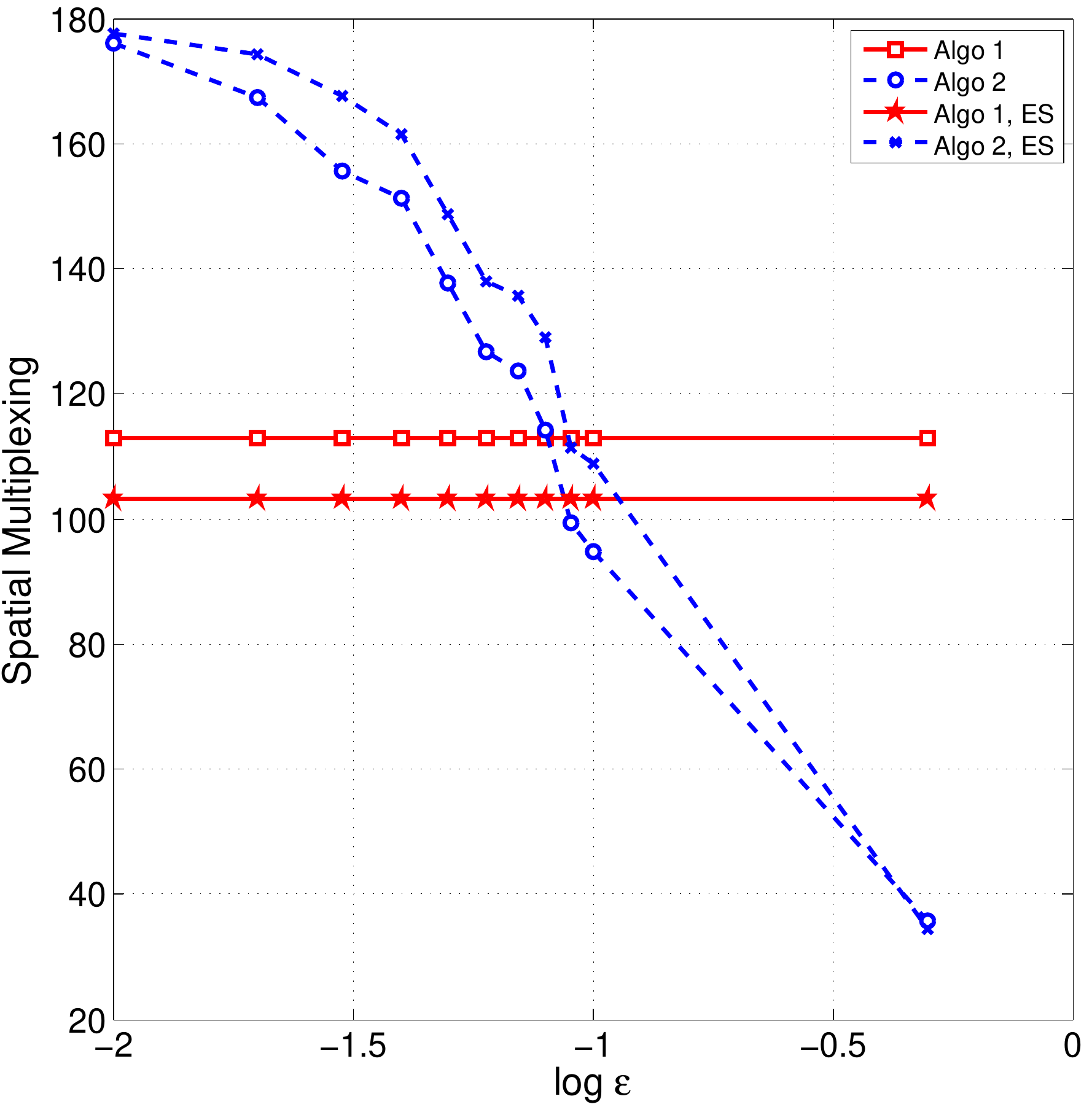}
  \label{fig:user-mux-mcs-5}
  }
  \caption{Comparison of Spatial Multiplexing versus $\log \epsilon$ with $G = 2$ and $G = 5$ user groups.
  Each user group has multiple scattering clusters, of which some are common to more than one group.}\label{fig:user-mux-mcs}
\end{figure}

Figures \ref{fig:rate-mux-mcs-2} and \ref{fig:rate-mux-mcs-5} show a comparison of the total achievable throughput for the different
algorithms as a function of SNR. {\em ``Algo 1''} refers to {\em Greedy Algorithm 1}, {\em ``Algo 2''} refers to {\em Greedy Algorithm 2} { and {\em ``ES''} refers to Exhaustive Search.}
We see that both algorithms give similar performance, with Algorithm 1 giving better performance than Algorithm 2 when the number of user
groups is 5. The average number of users simultaneously served, i.e., the spatial multiplexing, per time-frequency resource is
plotted in Figures \ref{fig:user-mux-mcs-2} and \ref{fig:user-mux-mcs-5}. Even though Algorithm 2 gives higher spatial multiplexing compared to Algorithm 1,
the presence of more groups reduces the beamforming gain and also creates additional inter-group interference (a result of non-perfect block diagonalization),
therefore, the gains due to spatial multiplexing are not fully realized. It is also noteworthy to observe the effect of $\epsilon$ as a tuning parameter.
A lower value of $\epsilon$ favors the selection of more groups (multiplexing) but in this case yields lower throughput because of
the smaller beamforming gain and higher inter-group interference. Instead, a higher value of $\epsilon$ sacrifices some spatial
dimensions but yields higher throughput in this case. { It is also noteworthy to point out that both the greedy user selection algorithms give good performance when compared with their exhaustive search counterparts, evidenced by Figures \ref{fig:rate-mux-mcs-5} and \ref{fig:user-mux-mcs-5}, for $G = 5$.\footnote{The fact that the spatial multiplexing of Algorithm 1 using exhaustive search may be less than what obtained by the greedy algorithm (as in Fig. \ref{fig:user-mux-mcs-5}) can be expected, since Algorithm 1 does not maximize the multiplexing gain.}}

\subsection{Covariance-based JSDM}
\setcounter{paragraph}{0}

We apply the covariance-based JSDM  scheme outlined in Section \ref{subsec:jsdm-after-us} to different scenarios,
and shall see that this scheme is particularly suited to directional channel models having a small number
of discrete MPCs.

\paragraph{User groups with multiple scattering clusters} \label{para-mcs-grp}

\begin{figure} [!h]
  \centering
  \subfigure[Sum Rate]{
  \includegraphics[width=7cm]{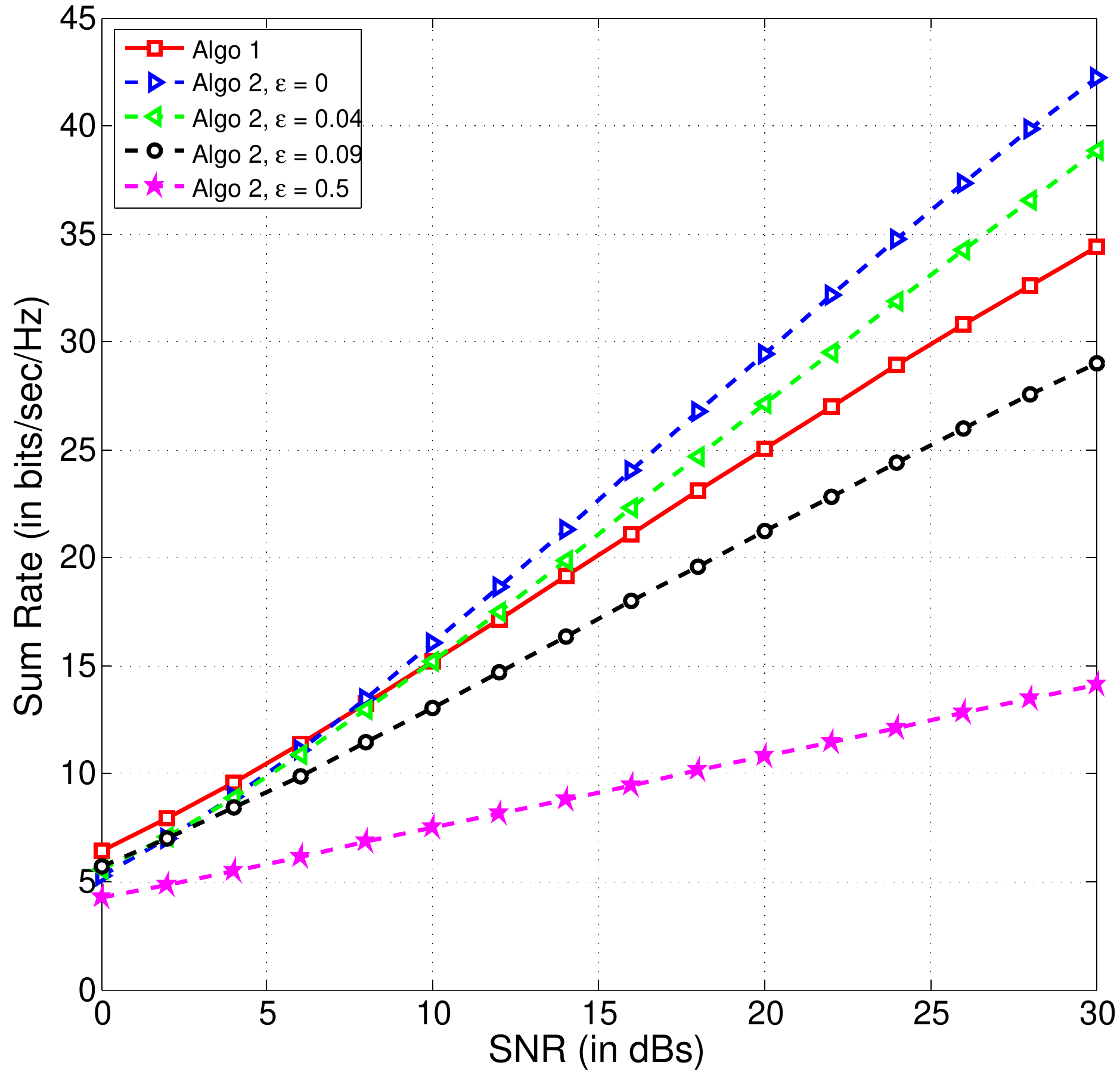}
  \label{fig:rate-no-mux-mcs-5}
  }
  \subfigure[Spatial Multiplexing]{
  \includegraphics[width=7cm]{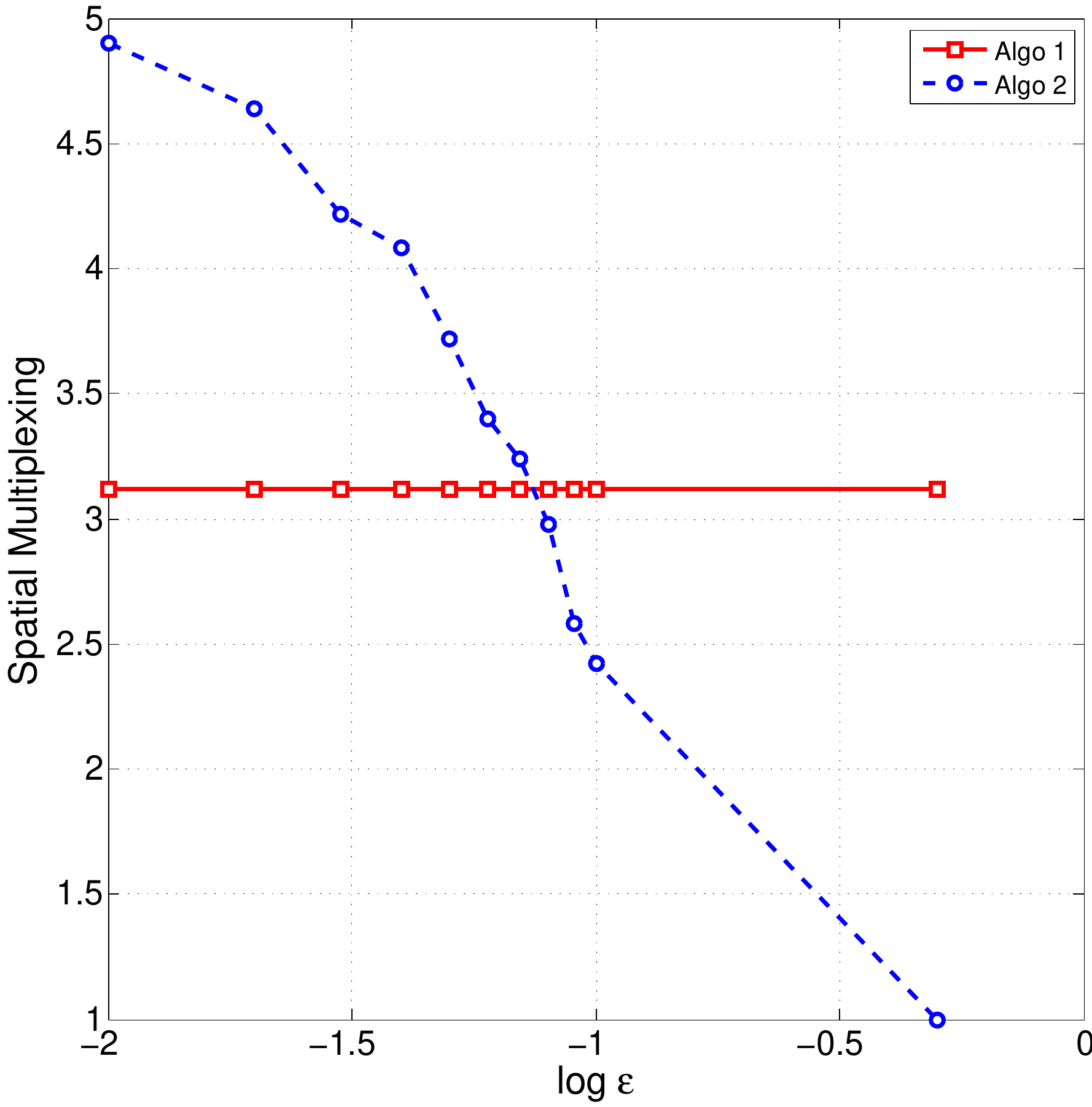}
  \label{fig:mux-gain-no-mux-mcs-5}
  }
  \caption{Comparison of sum spectral efficiency versus SNR and Spatial Multiplexing versus $\log \epsilon$ with $G = 2$ user groups and no spatial multiplexing. Each user has multiple scattering clusters.}\label{fig:no-mux-mcs-5}
\end{figure}

We consider the same setup as in Section \ref{subsec:jsdm-mux-mcs}.
As already remarked,  covariance-based JSDM serves only one user per group and does not require instantaneous CSIT
of the effective channels after pre-beamforming. Therefore, the precoder can be computed only from the second order statistics, eliminating the need for
explicit downlink training and simplifying the precoder design.
However, a price is paid in terms of achievable throughput, which is reduced considerably with respect to the full JSDM case.
Figure \ref{fig:rate-no-mux-mcs-5} shows the sum spectral efficiency as a function of SNR for the different user selection
algorithms and Figure \ref{fig:mux-gain-no-mux-mcs-5} shows the corresponding spatial multiplexing, when there are $G = 5$ groups.
Compared to Figures \ref{fig:rate-mux-mcs-5} and \ref{fig:user-mux-mcs-5}, there is a huge reduction in the achievable data rates and in
the spatial multiplexing.

\paragraph{Isolated Users with Multiple Scattering Clusters} \label{para-mcs}

\begin{figure} [!h]
  \centering
  \subfigure[Sum Rate]{
  \includegraphics[width=7cm]{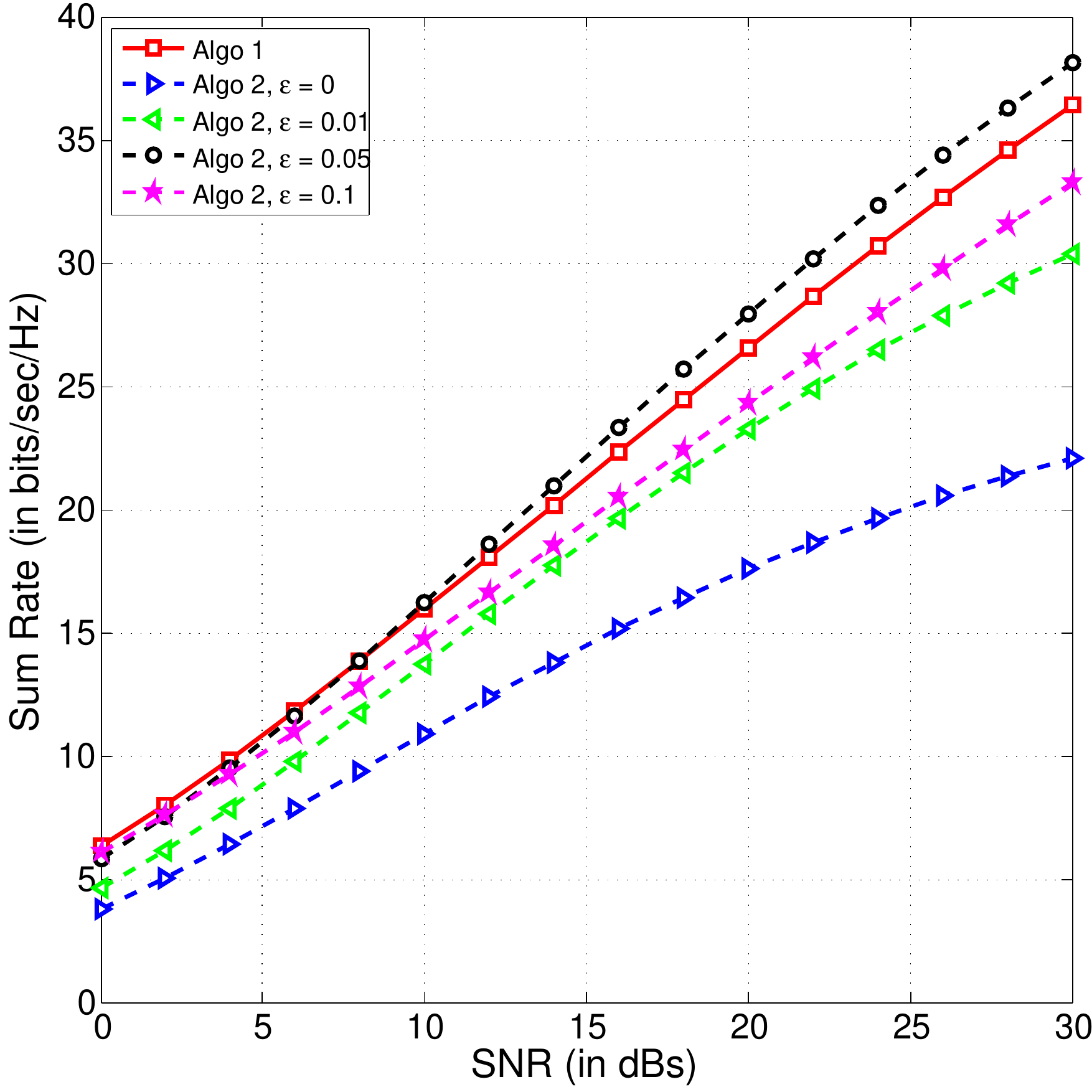}
  \label{fig:rate-no-mux-mcs}
  }
  \subfigure[Spatial Multiplexing]{
  \includegraphics[width=7cm]{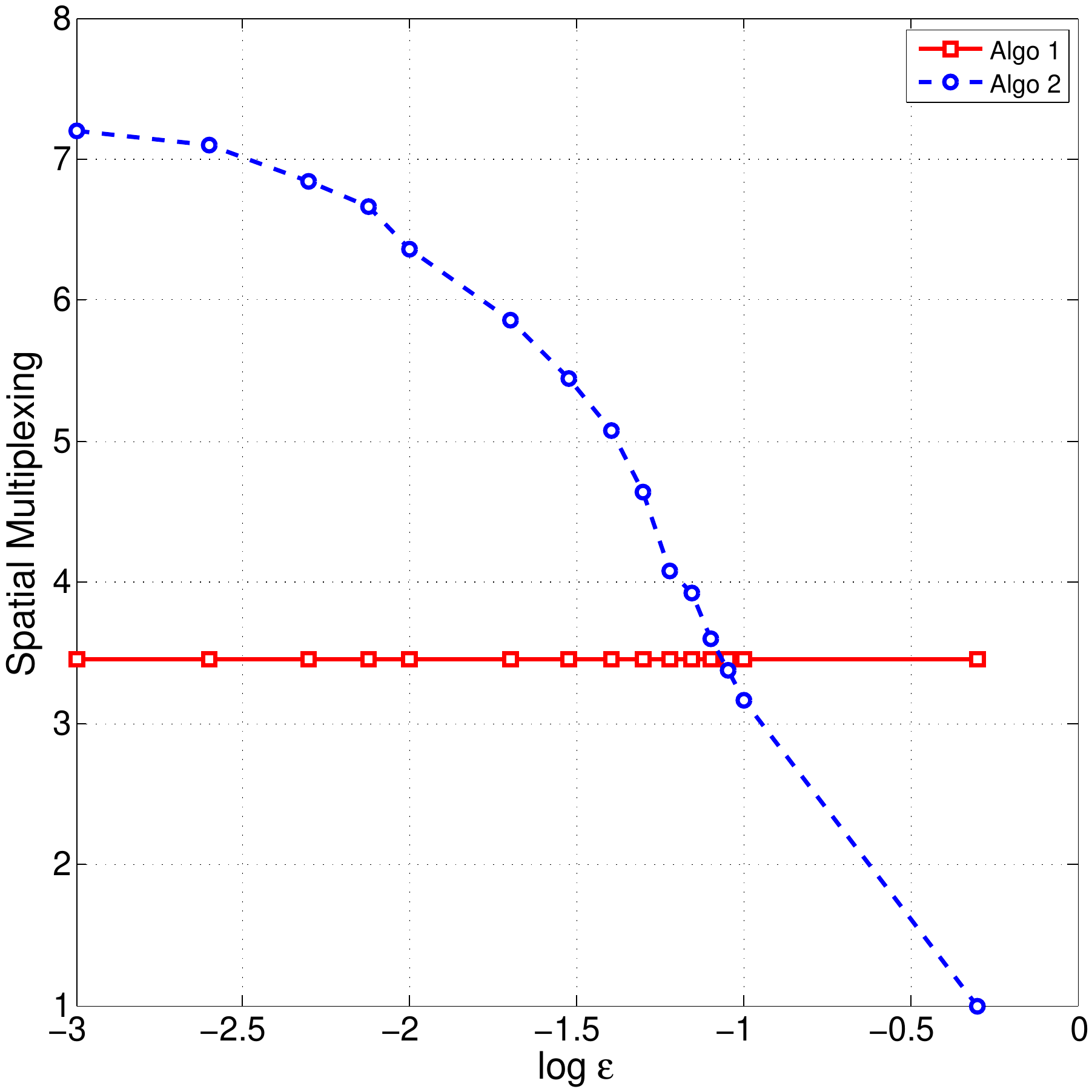}
  \label{fig:mux-gain-no-mux-mcs}
  }
  \caption{Comparison of sum spectral efficiency versus SNR and Spatial Multiplexing versus $\log \epsilon$ with $K = 20$ users. Each user has multiple scattering clusters.}\label{fig:user-no-mux-mcs}
\end{figure}

Here, we consider multiple scattering clusters associated to each user, similar to Section \ref{subsec:jsdm-mux-mcs}.
We fix the number of users in the system to be $K = 20$, and associate an arbitrary number of disjoint scattering clusters to each user.
The maximum number of scattering clusters that a user can have is limited to $5$. We set $M = 400$ and obtain a set of scheduled users by running the algorithms of Section \ref{sec:usr-sel}. Figure \ref{fig:rate-no-mux-mcs} shows the sum spectral efficiency with varying SNR for this setup and
Figure \ref{fig:mux-gain-no-mux-mcs} shows the variation of spatial multiplexing with the tuning parameter $\epsilon$.
We observe a behavior similar to what was observed for the model used in \ref{subsec:jsdm-mux-mcs}, and the achievable throughput is reduced significantly due to no spatial multiplexing. Also interesting is the fact that even though there are a total of $K = 20$ users, only an average of $7$ users are served simultaneously,
implying that the presence of more users leads to more common scattering clusters, thereby limiting the total spatial multiplexing.
This result might give the wrong intuition that having a larger number of users does not necessarily increase the total system throughput.
However, this effect is due to the limitation of the covariance-based JSDM: if full JSDM is used, users spanning the same set of dimensions
can be grouped together and served using MU-MIMO spatial multiplexing based on the instantaneous CSIT.
Interestingly, we shall see next that covariance-based JSDM is indeed able to achieve high spatial multiplexing (that increases with the number of users, in the range $K \ll M$)
in the presence of highly directional channels with a small number of MPCs.

\paragraph{Ray-tracing Based Channels}

\begin{figure} [!h]
  \centering
  \subfigure[$K = 5,10,25$]{
  \includegraphics[width=7cm]{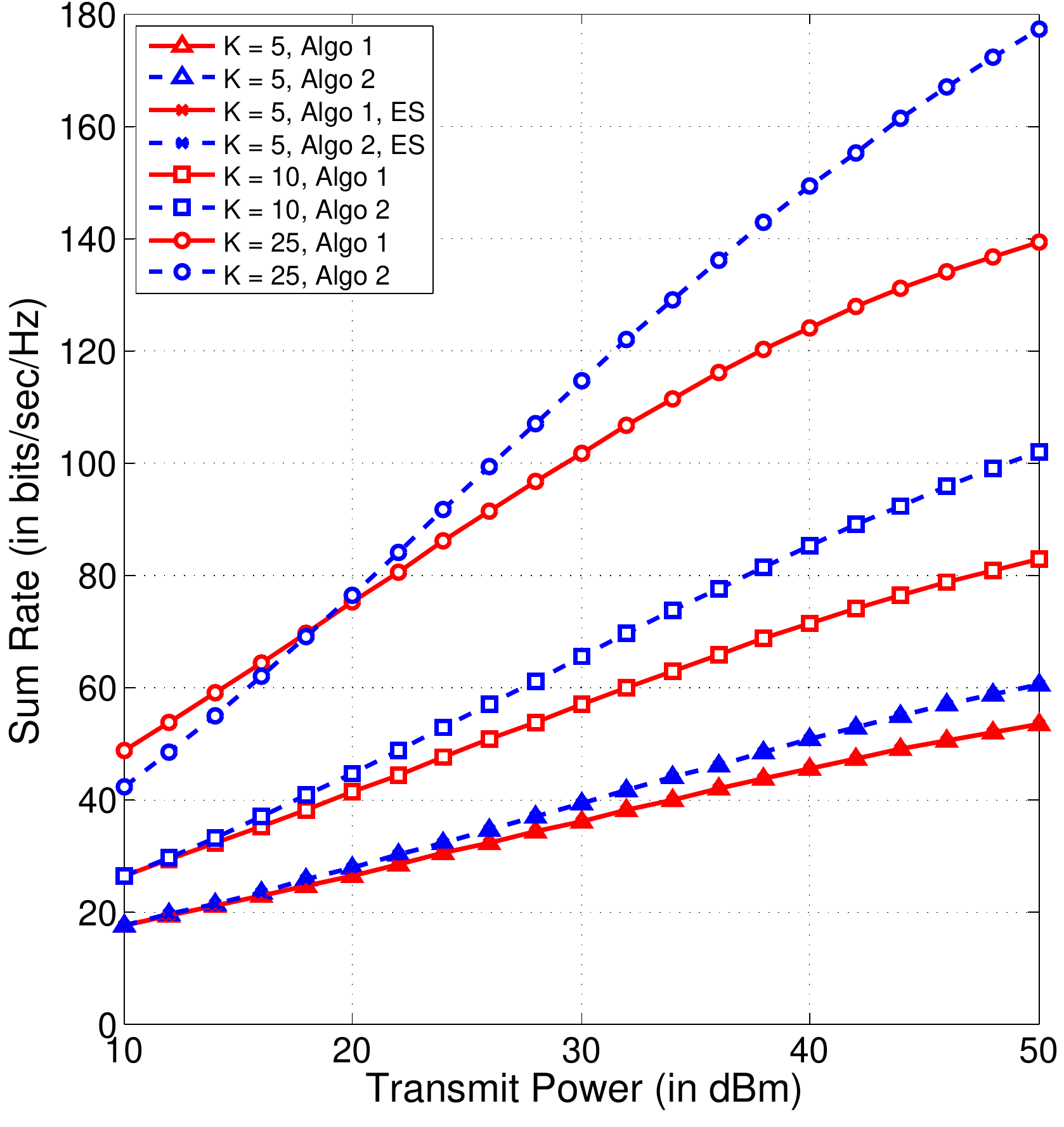}
  \label{fig:ddir-1}
  }
  \subfigure[$K = 45,60$]{
  \includegraphics[width=7cm]{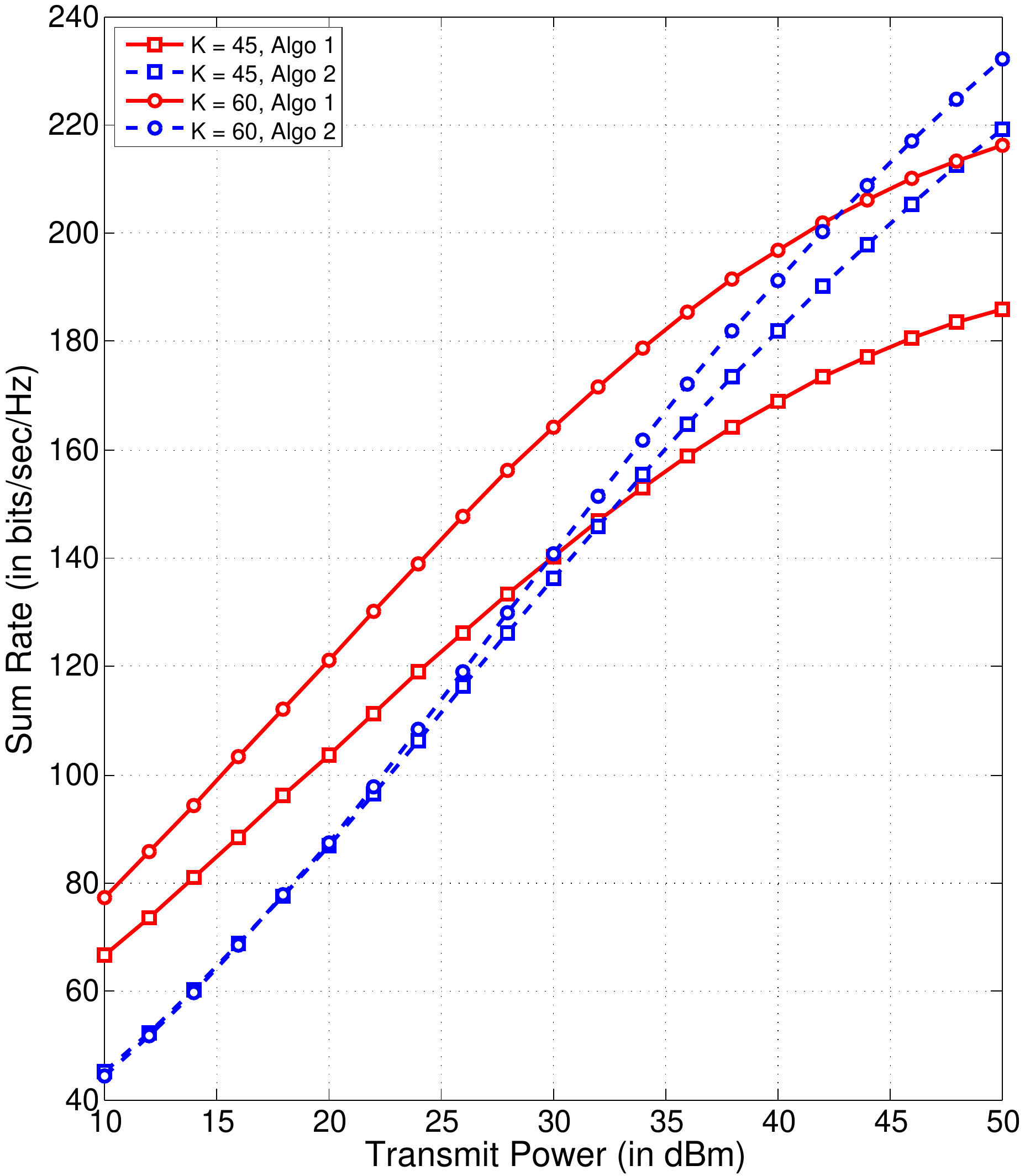}
  \label{fig:ddir-2}
  }
  \caption{Comparison of sum spectral efficiency versus transmit power with varying $K$ when the channel is modeled as a double directional impulse response.}\label{fig:ddir}
\end{figure}

\begin{figure} [!h]
  \centering
  \includegraphics[width=7cm]{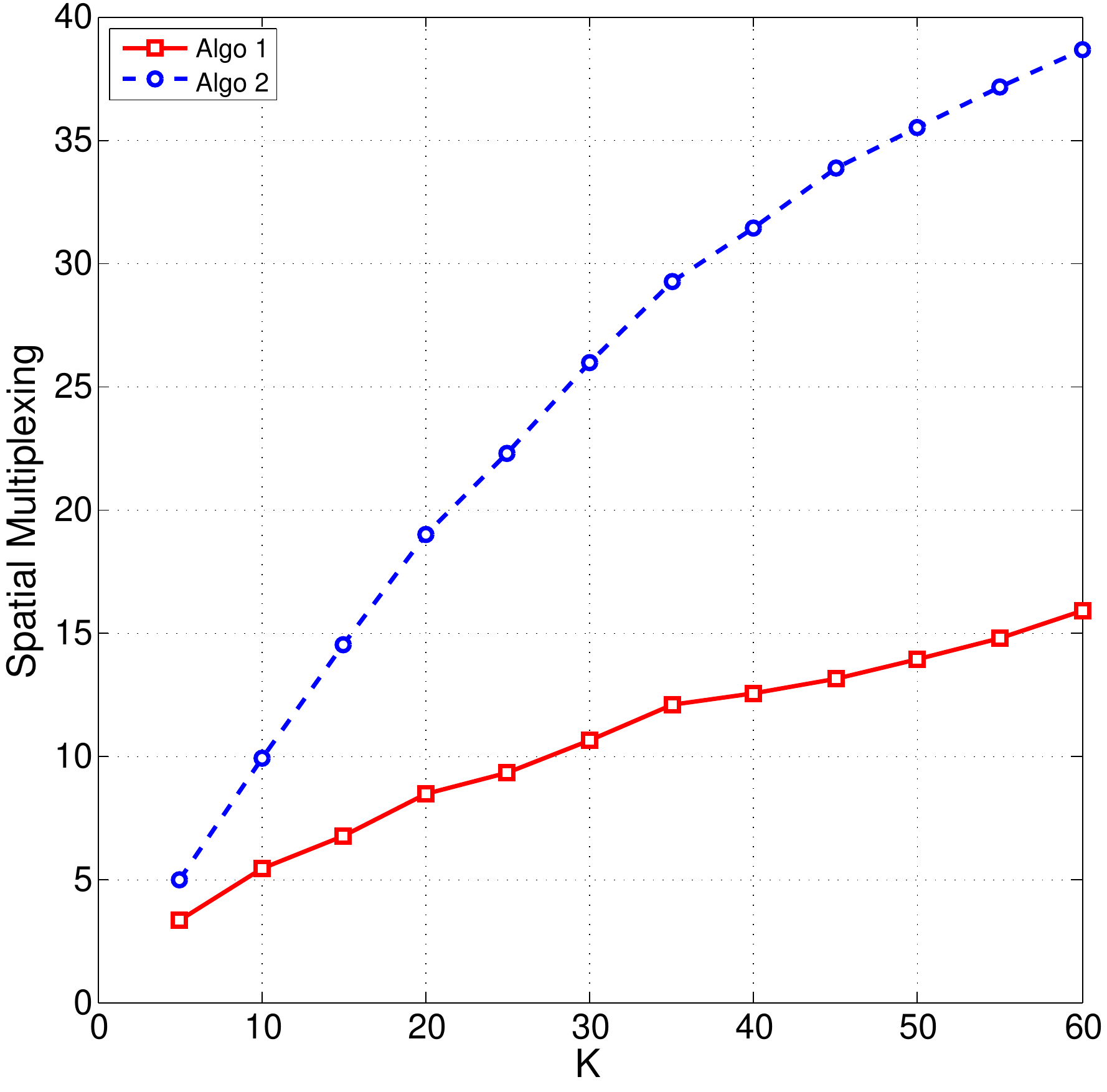}
  \caption{Comparison of Spatial Multiplexing versus Number of users when the channel is modeled as a double directional impulse response.}\label{fig:mux-gain}
\end{figure}

We next generate the channels according to (\ref{multipath}) by using parameters obtained from the ray-tracing simulation setup. The phases are
generated as $\phi_{kp} \sim {\rm Unif}[0,2\pi]$.
Since in this case the channel angular support is formed by a collection of disjoint ``angular frequency bins''
of the same size (see Section \ref{binsbins}), different user channels either do not overlap or overlap entirely on an integer number of bins.
Therefore, in algorithm 2 we can set $\epsilon = 0$. After obtaining the scheduled user set,
BD is performed to obtain the pre-beamformers.  Figure \ref{fig:ddir} shows the sum spectral efficiency versus transmit power (in dBm)
for various number of users with different algorithms. We vary the transmit power between $10$ dBm (10 mW) to $50$ dBm (100 W). The noise power is set to $-100$ dBm, corresponding to a $20$ MHz bandwidth. Here, we clearly see a tradeoff between {\em Orthogonalization} at low SNR and {\em Multiplexing} at high SNR. Also interesting is the fact that spatial multiplexing performs better with a small number of users than with a large number of users.
This is because there is a non-trivial tradeoff between Orthogonalization and Multiplexing. { With a lower complexity, greedy user selection performs well when compared with exhaustive search, as is clear from Figure \ref{fig:ddir-1} for $K = 5$.}
Contrary to what was observed in the case of multiple scattering clusters in Section \ref{para-mcs},
Figure \ref{fig:mux-gain} shows that we are able to recover the spatial multiplexing even with just covariance-based JSDM
when channels are highly directional and have a few MPCs, which characterize the channels obtained from ray tracing.

\paragraph{Measured Propagation Channels}

\begin{figure} [!h]
  \centering
  \includegraphics[width=7cm]{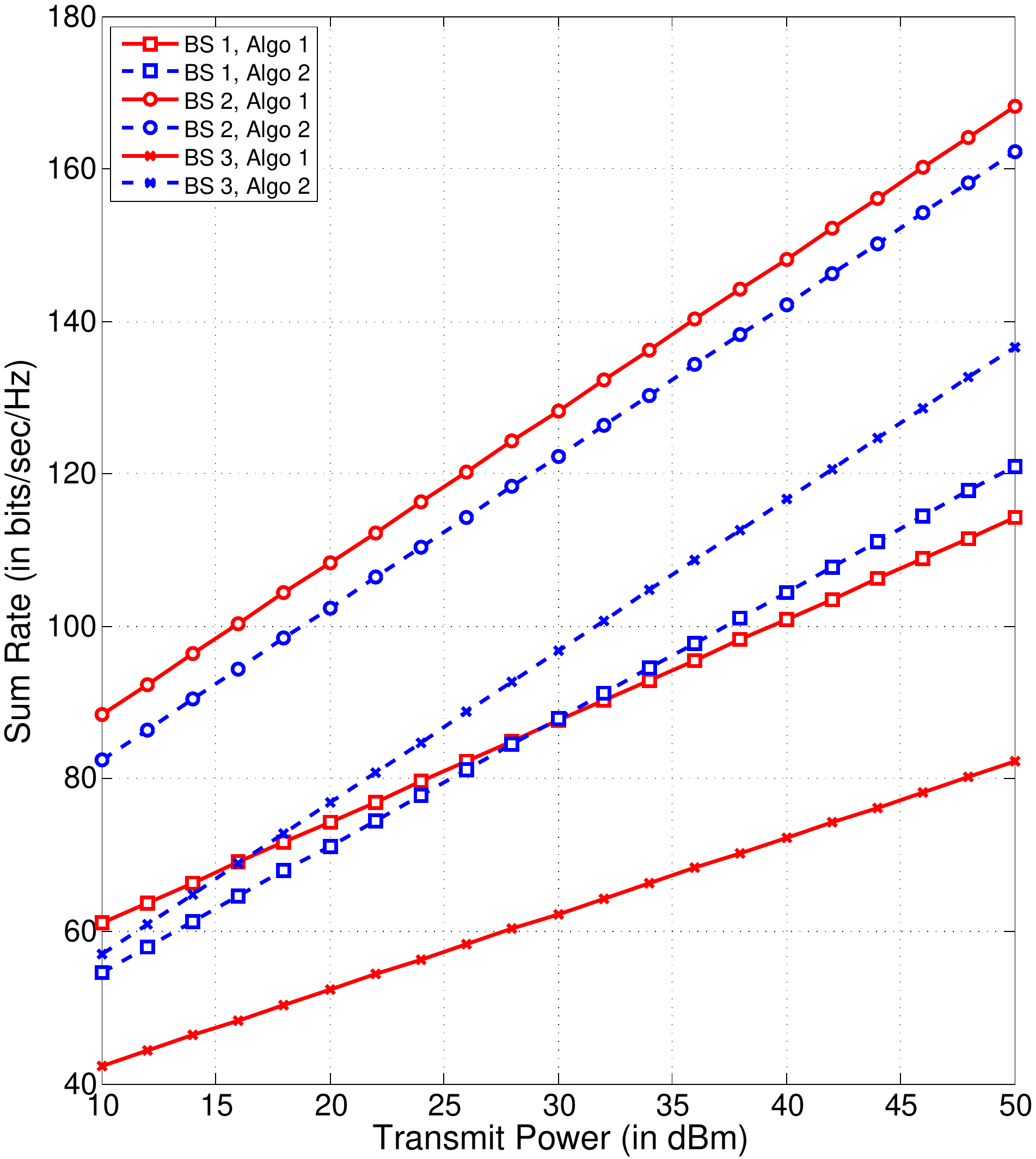}
  \caption{Comparison of sum spectral efficiency versus Transmit Power for different BS locations obtained from measured data.}\label{fig:md}
\end{figure}

Figure \ref{fig:md} shows the sum throughput versus SNR after running the user selection algorithms on the data obtained from measured propagation channels described in Section \ref{subsec:measurement}. There are a total of 3 BSs, and each BS has a set of 8 user locations, so we fix the number of users $K = 8$. We see that the algorithms perform differently depending on the scenario. For example, with BS 2, we achieve the same spatial multiplexing using both algorithms, while for BS 3, Algorithm 2 outperforms Algorithm 1 owing to huge spatial multiplexing. Overall, we observe that covariance-based JSDM along with proper user selection achieves very high throughput in actual propagation channels. However, one should also consider that the high spectral efficiencies are due to a single cell scenario and use of achievable rate expressions assuming Gaussian inputs. In reality, the input signal would be modulated by a finite dimensional constellation such as QAM, which would put a limit on the maximum achievable rate. Also, the noise floor was taken to be $-100$ dBm in our results, which is typical for a system operating at a bandwidth of 20 MHz under room temperature. Since inter-cell interference would create additional noise, this would reduce our received SNR too. Even taking into account all these imperfections, we would like to point out that in mm-Wave scenarios, the distances are short leading to smaller path losses and owing to the fact that we have a large antenna array at the BS, it is indeed possible to achieve high SNR with simple covariance based schemes, leading to high data rates.

{
\begin{rem}
Note that the proposed user selection algorithms are, in fact, independent of the channel model and use only the second order statistics of the user channels. However, these algorithms work well in certain kinds of channel environments such as those considered in the paper, and may not work well in other propagation environments. For example, if we have a few users with isotropic scattering, for which the energy is not concentrated in a particular angular direction but is distributed uniformly over the whole angular space, our selection algorithm will treat each of these users as a group on its own, and would either schedule one of these users alone, or multiple users with compatible directional channels. In terms of spatial multiplexing as well as reduced CSIT, our proposed algorithms become meaningful when most users in the network have channels with energy concentrated in a few directions. However, if we are in a propagation environment where most users have ``nearly'' isotropic channel directions, the scheme reduces to serving one user at a time, or a group of users based on instantaneous CSIT, as is the case in standard massive MIMO schemes.
\end{rem}}

\section{Conclusion} \label{sec:conc}

In this work we have considered the application of the JSDM approach to highly directional channels formed by a few discrete MPCs, or clusters
of multi-path components, typically arising in outdoor mm-Wave  communications.
In particular, when the user channels have partially overlapping eigenspaces, due to common scattering clusters or MPCs with similar angles of departure,
allocating users onto the BS array angular dimensions becomes a difficult optimization problem.
We formulate this problem in terms of a conflict graph, where each user is identified by the set of angular frequencies occupied by its
channel covariance spectrum, and users with overlapping angular frequencies are connected in the graph.
The user selection and angular dimension allocation can be formulated as integer programming problems, whose objective function
depends on what we wish to optimize. Here, we have proposed two such problems, driven by the physical insights gained by considering common scattering clusters. For the proposed integer programming problems, we have provided solutions via low complexity greedy
selection algorithms. Then, we have demonstrated the performance achieved by JSDM with the proposed algorithms
in some relevant scenarios, including channels generated by ray tracing in an outdoor campus environment and
channels obtained by an actual measurement campaign in an urban environment.

In general, JSDM with good user selection turns out to be an attractive technique for the implementation of
multiuser MIMO downlink in massive MIMO systems. The scheme can take advantage of highly directional channel statistics, as those arising in
mm-Wave frequencies. In particular, in a typical small-cell scenario where the number of users is significantly less than the number of base station antennas, and the
user channels are formed by a small number of discrete multi-path components, we have proposed a simple ``covariance-based'' JSDM scheme
that achieves remarkable spatial multiplexing while requiring only the knowledge of the channel's second-order statistics.
This scheme is particularly attractive since it does not require instantaneous CSIT feedback, and the channel covariances can be accurately
learned and tracked since they depend on the scattering environment, and are very slowly varying for nomadic users typical of
small cell networks.

\end{document}